\documentclass[journal,11pt,draftclsnofoot,onecolumn]{IEEEtran}

\usepackage{graphicx}

\usepackage[sort,compress]{cite}
\usepackage[usenames,dvipsnames]{color}
\usepackage{latexsym}
\usepackage{psfrag}
\usepackage{amsthm,amssymb,amscd,amssymb,amsmath,mathrsfs,mathbbold}
\usepackage[noend]{algpseudocode}

\theoremstyle{remark}
\newtheorem*{prf}{Proof}
\newtheorem{lem}{Lemma}
\newtheorem{thm}{Theorem}
\newtheorem{defn}{Definition}

\newtheorem{prop}{Proposition}

\newtheorem{exmp}{Example}
\newtheorem{cor}{Corollary}

\newtheorem{rem}{Remark}

\begin{document}
%
\title{ \LARGE{Average Consensus on General Strongly Connected Digraphs} }
%
%
%


\author{Kai~Cai
        ~and~Hideaki~Ishii
\thanks{The authors are with the Department of Computational Intelligence and Systems Science,
Tokyo Institute of Technology, 4259-J2-54, Nagatsuta-cho, Midori-ku,
Yokohama 226-8502, Japan.  Phone/Fax: +81-45-924-5371.  Emails:
caikai@sc.dis.titech.ac.jp, ishii@dis.titech.ac.jp.  This work was
supported in part by the Ministry of Education, Culture, Sports,
Science and Technology in Japan under Grant-in-Aid for Scientific
Research, No. 21760323.}
}

\maketitle

\begin{abstract}
We study the average consensus problem of multi-agent systems for
general network topologies with unidirectional information flow. We
propose two (linear) distributed algorithms, deterministic and
gossip, respectively for the cases where the inter-agent
communication is synchronous and asynchronous. Our contribution is
that in both cases, the developed algorithms guarantee state
averaging on arbitrary strongly connected digraphs; in particular,
this graphical condition does not require that the network be
balanced or symmetric, thereby extending many previous results in
the literature. The key novelty of our approach is to augment an
additional variable for each agent, called ``surplus'', whose
function is to locally record individual state updates.  For
convergence analysis, we employ graph-theoretic and nonnegative
matrix tools, with the eigenvalue perturbation theory playing a
crucial role.
\end{abstract}
%


\section{Introduction} \label{Sec1_Intro}


This paper presents a new approach to the design of distributed
algorithms for \emph{average consensus}: that is, a system of
networked agents reaches an agreement on the average value of their
initial states, through merely local interaction among peers.  The
approach enables multi-agent systems to achieve average consensus on
arbitrary strongly connected network topologies with
unidirectional information flow, where the state sum of the agents
need not stay put as time evolves.


There has been an extensive literature addressing
multi-agent consensus problems.  Many fundamental distributed
algorithms (developed in, e.g.,
\cite{BerTsi:89,JadLinMor:03,OlfMur:04,XiBo:04,RenBea:08}) are of
the \emph{synchronous} type: At an arbitrary time,
individual agents are assumed to sense and/or communicate with all
the neighbors, and then simultaneously execute their local
protocols. In particular, Olfati-Saber and Murray \cite{OlfMur:04}
studied algorithms of such type to achieve average consensus on
static digraphs, and justified that a \emph{balanced} and
\emph{strongly connected} topology is necessary and sufficient to
guarantee convergence.  More recently in \cite{BoGhPrSh:06}, Boyd et
al. proposed a compelling ``gossip'' algorithm, which provides an
asynchronous approach to treat average consensus. Specifically, the
algorithm assumes that at each time instant, exactly one agent wakes
up, contacts only one of its neighbors selected at random, and then
these two agents average out their states.  The graph model that the
algorithm bases is undirected (or symmetric), and average consensus
is ensured as long as the topology is \emph{connected}. Since then,
the gossip approach has been widely adopted
\cite{KaBaSr:07,CaFaFrZa:Automatica10,LavMur:TAC10} in tackling
average consensus on undirected graphs, with additional constraints
on quantized information flow; see also \cite{IshTem:10} for related
distributed computation problems in search engines.


In this paper, and its conference precursor \cite{CaiIshii_CDC:11},
we study the average consensus problem under both
synchronous and asynchronous setups, as in \cite{OlfMur:04} and
\cite{BoGhPrSh:06}.  In both cases, we propose a novel type of
(linear) distributed algorithms, which can be seen as extensions of
the corresponding algorithms in \cite{OlfMur:04} and
\cite{BoGhPrSh:06}; and we prove that these new algorithms guarantee
state averaging on arbitrary strongly connected digraphs, therefore
generalizing the graphical conditions derived in \cite{OlfMur:04}
and \cite{BoGhPrSh:06}.
We note that digraph models have been studied extensively in the consensus literature
\cite{OlfMur:04,XiBo:04,RenBea:08}, and considered to be generally more economical
for information exchange than their undirected counterpart (refer to \cite{OlfMur:04}
for more detailed motivation of using digraphs).
Our underlying (theoretic) interest in this paper is to generalize
the connectivity conditions on digraphs for average consensus.

The primary challenge of average consensus on arbitrary strongly
connected digraphs lies in that the state sum of agents need not be
preserved, thereby causing shifts in the average.  We note that
there are a few efforts in the literature having addressed this
issue. In \cite{FraGiuSea:09}, an auxiliary variable is
associated to each agent and a linear broadcast gossip
algorithm is proposed; however, the convergence of that algorithm is
not proved, and remarked to be difficult.
References \cite{KemDobGeh:03,BBTTV:10} also use extra variables, and
a nonlinear (division involved)
algorithm is designed and proved to achieve state averaging on
non-balanced digraphs.  The idea is based on computing the
stationary distribution for the Markov chain characterized by the
agent network, and is thus quite different from consensus-type
algorithms \cite{OlfMur:04,BoGhPrSh:06}. In \cite[Section~7.4]{BerTsi:89}, the load balancing
problem is tackled in which inter-processor communication is
asynchronous and with bounded delay. The underlying topology is
assumed undirected; owing to asynchronism and delay, however, the
total amount of loads at processors is not invariant.  A switched
linear algorithm is proposed to achieve load balancing in this
scenario, the rules of which rely on however bidirectional communication.
In addition, a different and interesting approach is presented in
\cite{GhaCor:11}: Given a general strongly connected digraph, find a
corresponding doubly stochastic matrix (which, when used as a distributed updating scheme,
guarantees state averaging \cite{XiBo:04}). An algorithm is designed to
achieve this goal by adding selfloop edges with proper weights to balance flow-in
and -out information.
Finally, time-varying state sum caused by packet loss or communication failure is considered in
\cite{PaBaEl:07,FagZam:09}, and the deviation from the initial average is analyzed.


We develop a new approach to handle the
problem that the state sum of agents need not be preserved.  Similar to
\cite{FraGiuSea:09}, we also augment an additional variable
for each agent, which we call ``surplus''; different from
\cite{KemDobGeh:03,BBTTV:10}, the function of surplus variables is
to record every state change of the associated agent. Thus, in
effect, these variables collectively maintain the information of the
average shift amount.\footnote{The method of augmenting
auxiliary variables is also found in
\cite{AyOrCo:09} and \cite{LiuAndCaoMor:09}, as \emph{predictors} estimating future
states and \emph{shift registers} storing past states respectively, in
order to accelerate consensus speed. How the predictors or registers are
used in these references is, however, very different from our
usage of surpluses.} Using this novel idea, our main contribution is the
design of linear algorithms (without switching) to achieve average consensus on
general strongly connected digraphs, in contrast with the types of algorithms designed in
\cite{KemDobGeh:03,BBTTV:10} and \cite[Section~7.4]{BerTsi:89}.
Also, linearity allows us to employ certain matrix tools in analysis,
which are very different from the proof methods used in \cite{KemDobGeh:03,BBTTV:10}
and \cite[Section~7.4]{BerTsi:89}.  Moreover, our technical contribution in this
paper is the demonstration of matrix perturbation tools (including eigenvalue
perturbation, optimal matching distance, and Bauer-Fike Theorem \cite{SeyMai:04,Bha:96,SteSun:90})
in analyzing convergence properties, which seems unexplored in the consensus literature.


Our idea of adding
surpluses is indeed a continuation of our own previous work in
\cite{CaiIsh:TAC10}, where the original
surplus-based approach is proposed to tackle \emph{quantized}
average consensus on general digraphs. There we developed a
quantized (thus nonlinear) averaging algorithm, and the convergence
analysis is based on finite Markov chains. By contrast, the
algorithms designed in the present paper are linear, and hence the
convergence can be characterized by the spectral properties of the
associated matrices.
On the other hand, our averaging algorithms differ also
from those basic ones \cite{OlfMur:04,BoGhPrSh:06} in that the
associated matrices contain negative entries. Consequently for our
analysis tools, besides the usual nonnegative matrix theory and algebraic
graph theory, the matrix perturbation theory is found instrumental.


The paper is organized as follows. Section~\ref{Sec2_ProFor}
formulates distributed average consensus problems in both
synchronous and asynchronous setups. Sections~\ref{Sec3_SynAlg} and
\ref{Sec4_GosAlg} present the respective solution algorithms, which
are rigorously proved to guarantee state averaging on general
strongly connected digraphs. Further, Section~\ref{Sec5_SpeGra}
explores certain special topologies that lead us to specialized
results, and Section~\ref{Sec6_Simul} provides a set of numerical
examples for demonstration. Finally, Section~\ref{Sec7_Concl} states
our
conclusions. 


\emph{Notation.} Let $\textbf{1}:=[1 \cdots 1]^T \in \mathbb{R}^n$
be the vector of all ones.  For a complex number $\lambda$, denote
its real part by Re$(\lambda)$, imaginary part by Im$(\lambda)$,
conjugate by $\bar{\lambda}$, and modulus by $|\lambda|$.  For a set
$\mathcal {S}$, denote its cardinality by card$(\mathcal {S})$.
Given a real number $x$, $\lfloor x \rfloor$ is the largest integer
smaller than or equal to $x$, and $\lceil x \rceil$ is the smallest
integer larger than or equal to $x$.  Given a matrix $M$, $|M|$
denotes its determinant; the spectrum $\sigma(M)$ is the set of its
eigenvalues; the spectral radius $\rho(M)$ is the maximum modulus of
its eigenvalues. In addition, $||\cdot||_2$ and $||\cdot||_\infty$
denote the $2$-norm and infinity norm of a vector/matrix.


\section{Problem Formulation} \label{Sec2_ProFor}


Given a network of $n$ ($>1$) agents, we model its interconnection
structure by a digraph $\mathcal {G} = (\mathcal {V}, \mathcal
{E})$: Each \emph{node} in $\mathcal {V} = \{ 1,...,n \}$ stands for
an agent, and each directed \emph{edge} $(j,i)$ in $\mathcal {E}
\subseteq \mathcal {V} \times \mathcal {V}$ denotes that agent $j$
communicates to agent $i$ (namely, the information flow is from $j$
to $i$). Selfloop edges are not allowed, i.e., $(i,i) \notin
\mathcal {E}$. In $\mathcal {G}$ a node $i$ is \emph{reachable} from
a node $j$ if there exists a path from $j$ to $i$ which respects the
direction of the edges. We say $\mathcal {G}$ is \emph{strongly
connected} if every node is reachable from every other node. A
\emph{closed strong component} of $\mathcal {G}$ is a maximal set of
nodes whose corresponding subdigraph is strongly connected and
closed (i.e., no node inside the subdigraph is reachable from any
node outside). Also a node $i$ is called \emph{globally reachable}
if every other node is reachable from $i$.

At time $k \in \mathbb{Z}_+$ (nonnegative integers) each node $i \in
\mathcal {V}$ has a scalar state $x_i(k) \in \mathbb{R}$; the
aggregate state is denoted by $x(k) = [x_1(k) \cdots x_n(k)]^T \in
\mathbb{R}^n$.  The \emph{average consensus problem} aims at
designing distributed algorithms, where individual nodes update
their states using only the local information of their neighboring
nodes in the digraph $\mathcal {G}$ such that all $x_i(k)$
eventually converge to the initial average $x_{a}:=\textbf{1}^T
x(0)/n$. To achieve state averaging on general digraphs, the main difficulty
is that the state sum $\textbf{1}^T x$ need not remain invariant,
which can result in losing track of the initial average $x_{a}$. To
deal with this problem, we propose associating to each node $i$ an
additional variable $s_i(k) \in \mathbb{R}$, called \emph{surplus};
write $s(k) = [s_1(k) \cdots s_n(k)]^T \in \mathbb{R}^n$ and set
$s(0)=0$.  The function of surplus is to locally record the state
changes of individual nodes such that $\textbf{1}^T (x(k)+s(k)) =
\textbf{1}^T x(0)$ for all time $k$; in other words, surplus keeps
the quantity $\textbf{1}^T (x+s)$ constant over time.


In the first part of this paper, we consider synchronous networks as
in \cite{OlfMur:04}: At each time, every node communicates with all
of its neighbors simultaneously, and then makes a corresponding
update.

\begin{defn}
A network of agents is said to achieve \emph{average consensus} if
for every initial condition $(x(0),s(0)=0)$, it holds that $(x(k),s(k)) \rightarrow
(x_a \textbf{1}, 0)$ as $k \rightarrow \infty$.
\end{defn}

\noindent \emph{Problem} 1. Design a distributed algorithm such that
agents achieve average consensus on general digraphs.

To solve this problem, we will propose in Section~\ref{Sec3_SynAlg}
a surplus-based distributed algorithm, under which we justify that
average consensus is achieved for general strongly connected digraphs.


In the second part, we consider the setup of asynchronous networks
as in \cite{BoGhPrSh:06}.  Specifically, communication among nodes
is by means of \emph{gossip}: At each time, exactly one edge $(j,i)
\in \mathcal {E}$ is activated at random, independently from all
earlier instants and with a time-invariant, strictly positive
probability $p_{ij} \in (0,1)$ such that $\sum_{(j,i) \in \mathcal
{E}} p_{ij} = 1.$  Along this activated edge, node $j$ sends its
state and surplus to node $i$, while node $i$ receives the
information and makes a corresponding update.

\begin{defn}
A network of agents is said to achieve


\noindent (i) \emph{mean-square average consensus} if for every
initial condition $(x(0),s(0)=0)$, it holds that \\ $E\left[||x(k)-x_{a}\textbf{1}
||_2^2\right] \rightarrow 0$ and $E\left[||s(k)||_2^2\right]
\rightarrow 0$
as $k \rightarrow \infty$;

\noindent (ii) \emph{almost sure average consensus} if for every
initial condition $(x(0),s(0)=0)$, it holds that \\ $(x(k),s(k)) \rightarrow
(x_{a}\textbf{1},0)$ as $k \rightarrow \infty$ with probability one.
\end{defn}

As defined, the mean-square convergence is concerned with the second
moments of the state and surplus processes, whereas the almost sure
convergence is with respect to the corresponding sample paths.  It
should be noted that in general there is no implication between
these two convergence notions (e.g., \cite[Section~7.2]{GriSti:01}).


\noindent \emph{Problem} 2. Design a distributed algorithm such that
agents achieve mean-square and/or almost sure average consensus
on general digraphs.

For this problem, we will propose in Section~\ref{Sec4_GosAlg} a
surplus-based gossip algorithm, under which we justify that both
mean-square and almost sure average consensus can be achieved for
general strongly connected digraphs.



\section{Averaging in Synchronous Networks} \label{Sec3_SynAlg}

This section solves Problem 1.  First we present a (discrete-time)
distributed algorithm based on surplus, which may be seen as an
extension of the standard consensus algorithms in the literature
\cite{BerTsi:89,JadLinMor:03,OlfMur:04,XiBo:04,RenBea:08}. Then we prove convergence to
average consensus for general strongly connected digraphs.

\subsection{Algorithm Description}

Consider a system of $n$ agents represented by a digraph $\mathcal
{G} = (\mathcal {V}, \mathcal {E})$.  For each node $i \in \mathcal
{V}$, let $\mathcal {N}_i^+:=\{j \in \mathcal {V}:(j,i)\in \mathcal
{E}\}$ denote the set of its ``in-neighbors'', and $\mathcal
{N}_i^-:=\{h \in \mathcal {V}:(i,h)\in \mathcal {E}\}$ the set of
its ``out-neighbors''.  Note that $\mathcal {N}_i^+ \neq \mathcal {N}_i^-$ in
general; and $i \notin \mathcal {N}_i^+$ or
$\mathcal {N}_i^-$, for selfloop edges do not exist.  There are three
operations that every node $i$ performs at time $k \in
\mathbb{Z}_+$.  First, node $i$ sends its state information $x_i(k)$
and weighted surplus $b_{ih} s_i(k)$ to each out-neighbor $h \in
\mathcal {N}_i^-$; here the \emph{sending weight} $b_{ih}$ is
assumed to satisfy that $b_{ih}\in (0,1)$ if $h \in \mathcal
{N}_i^-$, $b_{ih}=0$ if $h\in \mathcal {V}-\mathcal {N}_i^-$, and
$\sum_{h \in \mathcal {N}_i^-} b_{ih}<1$. Second, node $i$ receives
state information $x_j(k)$ and weighted surplus $b_{ji} s_j(k)$ from
each in-neighbor $j \in \mathcal {N}_i^+$. Third, node $i$ updates
its own state $x_i(k)$ and surplus $s_i(k)$ as follows:
\begin{align}\label{eq:DetAlg_state}
x_i(k+1) &= x_i(k) + \sum_{j \in \mathcal {N}_i^+}
         a_{ij}(x_j(k)-x_i(k)) + \epsilon s_i(k), \\\label{eq:DetAlg_surplus}
s_i(k+1) &= \Big( ( 1-\sum_{h \in \mathcal {N}_i^-} b_{ih} ) s_i(k)
         + \sum_{j \in \mathcal {N}_i^+} b_{ji}s_j(k) \Big) -
         \Big(x_i(k+1)-x_i(k)\Big),
\end{align}
where the \emph{updating weight} $a_{ij}$ is assumed to satisfy that
$a_{ij}\in (0,1)$ if $j \in \mathcal {N}_i^+$, $a_{ij}=0$ if $j\in
\mathcal {V}-\mathcal {N}_i^+$, and $\sum_{j \in \mathcal
{N}_i^+}a_{ij}<1$; in addition, the parameter $\epsilon$ is a
positive number which specifies the amount of surplus used to update
the state.

We discuss the implementation of the above protocol in applications
like sensor networks. Let $\mathcal {G} = (\mathcal {V}, \mathcal
{E})$ represent a network sensor nodes.  Our protocol deals
particularly with scenarios where (i) sensors have different
communication ranges owing possibly to distinct types or power
supplies; (ii) communication is by means of broadcasting (e.g.,
\cite{FraGiuSea:09}) which again might have different ranges; and
(iii) strategy of random geographic routing is used for efficient
and robust node value aggregation in one direction
\cite{KemDobGeh:03,BBTTV:10}. In these scenarios, information flow
among sensors is typically directed.  A concrete example is using
sensor networks for monitoring geological areas (e.g., volcanic
activities), where sensors are fixed at certain locations. At the
time of setting them up, the sensors may be given different
transmission power for saving energy (such sensors must run for a
long time) or owing to geological reasons. Once the power is fixed,
the neighbors (and their IDs) can be known to each sensor. Thus,
digraphs can arise in \emph{static} sensor networks where the
neighbors can be fixed and known.  To implement states and
surpluses, we see from (\ref{eq:DetAlg_state}),
(\ref{eq:DetAlg_surplus}) that they are ordinary variables locally
stored, updated, and exchanged; thus they may be implemented by
allocating memories in sensors.
For the parameter $\epsilon$, we will see that it plays a
crucial role in the convergence of our algorithm; however,
$\epsilon$ must be chosen sufficiently small, and a valid bound for
its value involves non-local information of the digraph. The latter
constraint (in bounding a parameter) is often found in consensus
algorithms involving more than one variable
\cite{RenBea:08,LDCH:10,LFXZ:11}. One may overcome this by computing
a valid bound offline, and notify that $\epsilon$ value to every
node.

Now let the \emph{adjacency matrix} $A$ of the digraph $\mathcal {G}$ be
given by $A:=[a_{ij}]\in \mathbb{R}^{n \times n}$, where the entries
are the updating weights. Then the \emph{Laplacian matrix} $L$ is
defined as $L:=D-A$, where $D=\mbox{diag}(d_1, \ldots, d_n)$ with
$d_i=\sum_{j=1}^n a_{ij}$. Thus $L$ has nonnegative diagonal
entries, nonpositive off-diagonal entries, and zero row sums.
Then the matrix $I-L$ is nonnegative (by $\sum_{j \in \mathcal
{N}_i^+}a_{ij}<1$), and every row sums up to one; namely $I-L$ is
\emph{row stochastic}. Also let $B:=[b_{ih}]^T \in \mathbb{R}^{n
\times n}$, where the entries are the sending weights (note that the
transpose in the notation is needed because $h \in \mathcal {N}_i^-$
for $b_{ih}$). Define the matrix $S:=(I-\tilde{D})+B$, where
$\tilde{D}=\mbox{diag}(\tilde{d}_1, \ldots, \tilde{d}_n)$ with
$\tilde{d}_i=\sum_{h=1}^n b_{ih}$. Then $S$ is nonnegative (by
$\sum_{h \in \mathcal {N}_i^-} b_{ih}<1$), and every column sums up
to one; i.e., $S$ is \emph{column stochastic}. As can be observed
from (\ref{eq:DetAlg_surplus}), the matrix $S$ captures the part of
update induced by sending and receiving surplus.

With the above matrices, the iterations (\ref{eq:DetAlg_state}) and
(\ref{eq:DetAlg_surplus}) can be written in a matrix form as
\begin{align} \label{eq:DetAlg}
\begin{bmatrix}
  x(k+1)\\
  s(k+1)
\end{bmatrix}=
M
\begin{bmatrix}
  x(k)\\
  s(k)
\end{bmatrix},
\ \ \ \mbox{where } M:=
\begin{bmatrix}
  I-L & \epsilon I\\
  L & S-\epsilon I
\end{bmatrix} \in \mathbb{R}^{2n \times 2n}.
\end{align}
Notice that (i) the matrix $M$ has negative entries due to the
presence of the Laplacian matrix $L$ in the $(2,1)$-block; (ii) the
column sums of $M$ are equal to one, which implies that the quantity
$x(k)+s(k)$ is a constant for all $k \geq 0$; and (iii) the state
evolution specified by the $(1,1)$-block of $M$, i.e.,
\begin{align} \label{eq:stand_consen}
x(k+1) = (I-L) x(k),
\end{align}
is that of the \emph{standard} consensus algorithm studied in the
literature (e.g., \cite{BerTsi:89,OlfMur:04,XiBo:04}).   We will
henceforth refer to (\ref{eq:DetAlg}) as the deterministic
algorithm, and analyze its convergence properties in the next
subsection.

%
%

\begin{figure}[!t]
  \centering
  \includegraphics[width=0.5\textwidth]{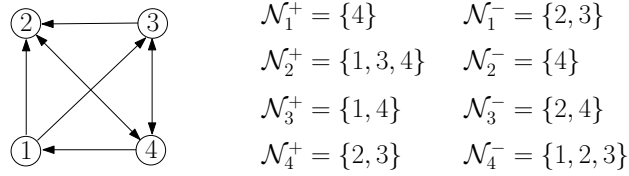}
  \vspace{-0.3cm}
  \caption{Illustrating example of four agents: communication topology and neighbor sets.}
  \label{fig:Ex_4nodes}
\end{figure}

\begin{exmp} \label{ex:detalg}
For an illustration of the algorithm (\ref{eq:DetAlg}), consider a
network of four nodes with neighbor sets shown in
Fig.~\ref{fig:Ex_4nodes}. Fixing $i\in[1,4]$, let
$a_{ij}=1/\left(\mbox{card}(\mathcal {N}_i^+)+1\right)$ for every $j
\in \mathcal {N}_i^+$ and $b_{ih}=1/\left(\mbox{card}(\mathcal
{N}_i^-)+1\right)$ for every $h \in \mathcal {N}_i^-$. Then the
matrix $M$ of this example is given by {\small
\begin{align*}
M=\left[
\begin{array}{cccc|cccc}
1/2 & 0 & 0 & 1/2 & \epsilon & 0 & 0 & 0 \\
1/4 & 1/4 & 1/4 & 1/4 & 0 & \epsilon & 0 & 0 \\
1/3 & 0 & 1/3 & 1/3 & 0 & 0 & \epsilon & 0 \\
0 & 1/3 & 1/3 & 1/3 & 0 & 0 & 0 & \epsilon \\
\hline 1/2 & 0 & 0 & -1/2 & 1/3-\epsilon & 0 & 0 & 1/4 \\
-1/4 & 3/4 & -1/4 & -1/4 & 1/3 & 1/2-\epsilon & 1/3 & 1/4 \\
-1/3 & 0 & 2/3 & -1/3 & 1/3 & 0 & 1/3-\epsilon & 1/4 \\
0 & -1/3 & -1/3 & 2/3 & 0 & 1/2 & 1/3 & 1/4-\epsilon
\end{array}\right].
\end{align*}
}
We see that $M$ has negative entries, and every column sums up to
one.
\end{exmp}




\subsection{Convergence Result}

The following is a graphical characterization for the deterministic algorithm (\ref{eq:DetAlg}) to
achieve average consensus. The proof is deferred to Section~\ref{pf1}.

\begin{thm} \label{thm:DetAlg}
Using the deterministic algorithm (\ref{eq:DetAlg}) with the
parameter $\epsilon>0$ sufficiently small, the agents achieve
average consensus if and only if the digraph $\mathcal {G}$ is
strongly connected.
\end{thm}

Without augmenting surplus variables, it is well known
\cite{OlfMur:04} that a necessary and sufficient graphical condition
for state averaging is that the digraph $\mathcal {G}$ is both
strongly connected and \emph{balanced}\footnote{A digraph $\mathcal
{G}$ with its adjacency matrix $A=[a_{ij}]$ is balanced if
$\sum_{j=1}^n a_{ij}=\sum_{j=1}^n a_{ji}$ for all $i$. Equivalently,
the system matrix $I-L$ of the standard consensus algorithm
(\ref{eq:stand_consen}) is both row and column stochastic
\cite{OlfMur:04,XiBo:04}.}.  A balanced structure can be
restrictive because when all the weights $a_{ij}$ are identical, it
requires the number of incoming and outgoing edges at each node in
the digraph to be the same. By contrast, our algorithm
(\ref{eq:DetAlg}) ensures average consensus
for arbitrary strongly connected digraphs (including those non-balanced).

A surplus-based averaging algorithm was initially proposed in
\cite{CaiIsh:TAC10} for a quantized consensus
problem. It guarantees that the integer-valued states converge to
either $\lfloor x_a \rfloor$ or $\lceil x_a \rceil$; however, the steady-state surpluses are nonzero in
general. There, the set of states and surpluses is finite, and thus
arguments of finite Markov chain type are employed in the proof.
Distinct from \cite{CaiIsh:TAC10},
with the algorithm (\ref{eq:DetAlg}) the states converge to the
exact average $x_a$ and the steady-state surpluses are zero.
Moreover, since the algorithm (\ref{eq:DetAlg}) is linear, its
convergence can be analyzed using tools from matrix theory, as
detailed below. This last linearity point is also in contrast with the division
involved algorithm designed in \cite{KemDobGeh:03,BBTTV:10}.

The choice of the parameter $\epsilon$ depends on the graph structure and
the number of agents.  In the following, we present an upper bound on $\epsilon$
for general networks.

\begin{prop} \label{prop:perturb_bound} Suppose that the digraph $\mathcal {G}$
is strongly connected.  The deterministic algorithm
(\ref{eq:DetAlg}) achieves average consensus if the parameter
$\epsilon$ satisfies $\epsilon \in (0,\bar{\epsilon}^{(d)})$, where
\begin{align} \label{eq:perturb_bound}
\bar{\epsilon}^{(d)} := \frac{1}{(20+8n)^n} \ (1-|\lambda_3|)^n,\ \
\mbox{with }\lambda_3 \mbox{ the third largest eigenvalue of $M$ by setting $\epsilon=0$}.
\end{align}
\end{prop}

The proof of Proposition~\ref{prop:perturb_bound} is presented in Section~\ref{pf2},  which
employs a fact from matrix perturbation theory (e.g., \cite{Bha:96,SteSun:90}) relating
the size of $\epsilon$ to the distance between perturbed and unperturbed eigenvalues.
Also, we will stress that this proof is based on that of Theorem~\ref{thm:DetAlg}.
The above bound $\bar{\epsilon}^{(d)}$ ensures
average consensus for arbitrary strongly connected topologies.  Due
to the power $n$, however, the bound is rather conservative.  This
power is unavoidable for any perturbation bound result with respect
to general matrices, as is well known in matrix perturbation
literature \cite{Bha:96,SteSun:90}.  In Section~\ref{Sec5_SpeGra},
we will exploit structures of some special topologies, which yield
less conservative bounds on $\epsilon$.  Also, we see that the bound
in (\ref{eq:perturb_bound}) involves $\lambda_3$, the second largest
eigenvalue of either $I-L$ or $S$ (matrix $M$ is block-diagonal when $\epsilon=0$).
This infers that, in order to
bound $\epsilon$, we need to know the structure of the
agent network. Such a requirement when bounding some parameters in
consensus algorithms, unfortunately, seems to be not unusual \cite{RenBea:08,LDCH:10,LFXZ:11}.

\subsection{Proof of Theorem~\ref{thm:DetAlg}} \label{pf1}

We present the proof of Theorem~\ref{thm:DetAlg}. First, we
state a necessary and sufficient condition for average consensus in
terms of the spectrum of the matrix $M$.

\begin{prop} \label{prop:Mean_AveCon} The deterministic algorithm
(\ref{eq:DetAlg}) achieves average consensus if and only if $1$ is a
simple eigenvalue of $M$, and all other eigenvalues have moduli
smaller than one.
\end{prop}
\begin{prf}
(Sufficiency) Since every column of $M$ sums up to one, $1$ is an
eigenvalue of $M$ and $[\textbf{1}^T \ \textbf{1}^T]^T$ is a
corresponding left eigenvector.  Note also that $M[\textbf{1}^T \
0]^T=[\textbf{1}^T \ 0]^T$;
so $[\textbf{1}^T \ 0]^T \in \mathbb{R}^{2n}$ is a right eigenvector
corresponding to the eigenvalue $1$. Write $M$ in Jordan canonical
form as
\begin{align*}
M = V J V^{-1} =\begin{bmatrix}
  y_1 & \cdots & y_{2n}
\end{bmatrix}
\begin{bmatrix}
  1 & 0 \\
  0 & J'
\end{bmatrix}
\begin{bmatrix}
  z_1^T \\ \vdots \\ z_{2n}^T
\end{bmatrix},
\end{align*}
where $y_i,\ z_i \in \mathbb{C}^{2n}$, $i\in [1,2n]$, are
respectively the (generalized) right and left eigenvectors of $M$;
and $J'\in \mathbb{C}^{(2n-1) \times (2n-1)}$ contains the Jordan
block matrices corresponding to those eigenvalues with moduli
smaller than one. For the eigenvalue $1$ choose $y_1=[\textbf{1}^T \
0]^T$ and $z_1=(1/n)[\textbf{1}^T \ \textbf{1}^T]^T$; thus $z_1^T
y_1 = 1$. Now the $k$th power of $M$ is
\begin{align*}
M^k = V J^k V^{-1}
    = V
\begin{bmatrix}
  1 & 0 \\
  0 & (J')^k
\end{bmatrix} V^{-1} \rightarrow y_1 z_1^T = \begin{bmatrix}
  \frac{1}{n} \textbf{1} \textbf{1}^T & \frac{1}{n} \textbf{1} \textbf{1}^T \\
  0 & 0
\end{bmatrix},\ \ \ \mbox{as $k \rightarrow \infty$.}
\end{align*}
Therefore
\begin{align*}
\begin{bmatrix}
  x(k)\\
  s(k)
\end{bmatrix}
=M^k
\begin{bmatrix}
  x(0)\\
  s(0)
\end{bmatrix}
\rightarrow \begin{bmatrix}
  \frac{1}{n} \textbf{1} \textbf{1}^T & \frac{1}{n} \textbf{1} \textbf{1}^T \\
  0 & 0
\end{bmatrix}
\begin{bmatrix}
  x(0)\\
  s(0)
\end{bmatrix} =
\begin{bmatrix}
  \frac{1}{n} \textbf{1} \textbf{1}^T x(0)\\
  0
\end{bmatrix} =
\begin{bmatrix}
  x_a \textbf{1}\\
  0
\end{bmatrix},\ \ \ \mbox{as $k \rightarrow \infty$.}
\end{align*}

(Necessity) First we claim that the eigenvalue $1$ of $M$ is always
simple.  Suppose on the contrary that the algebraic multiplicity of
$1$ equals two.  The corresponding geometric multiplicity, however,
equals one; this is checked by verifying rank$(M-I)=2n-1$. Thus
there exists a generalized right eigenvector $u=[u_1^T \ u_2^T]^T
\in \mathbb{R}^{2n}$ such that $(M-I)^2 u = 0$, and $(M-I) u$ is a
right eigenvector with respect to the eigenvalue $1$. Since
$[\textbf{1}^T \ 0]^T$ is also a right eigenvector corresponding to
the eigenvalue $1$, it must hold:
\begin{align*}
& (M-I) u = c [\textbf{1}^T \ 0]^T, \ \ \ \mbox{for some scalar $c \neq 0$}\\
\Rightarrow \ \ &
\begin{bmatrix}
  -L & \epsilon I \\
  L & S-I-\epsilon I
\end{bmatrix}
\begin{bmatrix}
  u_1 \\ u_2
\end{bmatrix} = c
\begin{bmatrix}
  \textbf{1} \\ 0
\end{bmatrix} \\
\Rightarrow \ \ & \left\{
  \begin{array}{ll}
    -L u_1 + \epsilon u_2 = c \textbf{1} \\
    L u_1 + (S-I)u_2  - \epsilon u_2
    =0
  \end{array}
\right. \\
\Rightarrow \ \ &(S-I)u_2 = c \textbf{1}.
\end{align*}
One may verify that rank$(S-I)=n-1$ but rank$(\left[S-I \ \ c
\textbf{1}\right])=n$.  Hence there is no solution for $u_2$, which
in turn implies that the generalized right eigenvector $u$ cannot
exist. This proves our claim.

Now suppose that there is an eigenvalue $\lambda$ of $M$ such that
$\lambda \neq 1$ and $|\lambda| \geq 1$. But this immediately
implies that $\lim_{k \rightarrow \infty} M^k$ does not exist
\cite{XiBo:04}.  Therefore, average consensus cannot be achieved.
\hfill $\blacksquare$
\end{prf}

Next, we introduce an important result from matrix perturbation
theory (e.g., \cite[Chapter~2]{SeyMai:04}), which is found crucial in
analyzing the spectral properties of the matrix $M$ in
(\ref{eq:DetAlg}). The proof of this result can be found in
\cite[Sections~2.8 and 2.10]{SeyMai:04}. An eigenvalue of a matrix
is said \emph{semi-simple} if its algebraic multiplicity is equal to
its geometric multiplicity.

\begin{lem} \label{lem:eigpur}
Consider an $n \times n$ matrix $W(\epsilon)$ which depends smoothly
on a real parameter $\epsilon \geq 0$. Fix $l \in [1,n]$; let
$\lambda_1=\cdots=\lambda_l$ be a semi-simple eigenvalue of $W(0)$,
with (linearly independent) right eigenvectors $y_1,\ldots,y_l$ and
(linearly independent) left eigenvectors $z_1,\ldots,z_l$ such that
\begin{align*}
\begin{bmatrix}
  z_1^T \\ \vdots \\ z_l^T
\end{bmatrix}
\begin{bmatrix}
  y_1 & \cdots & y_l
\end{bmatrix} = I.
\end{align*}
Consider a small $\epsilon>0$, and denote by $\lambda_i(\epsilon)$
the eigenvalues of $W(\epsilon)$ corresponding to $\lambda_i$, $i
\in [1,l]$. Then the derivatives $d\lambda_i(\epsilon)/d\epsilon |_{\epsilon=0}$
exist, and they are the eigenvalues of the following $l \times l$
matrix:
\begin{align} \label{matrix:eigpur}
\begin{bmatrix}
  z_1^T \dot{W} y_1  & \cdots & z_1^T \dot{W} y_l \\
  \vdots             &        & \vdots \\
  z_l^T \dot{W} y_1  & \cdots & z_l^T \dot{W} y_l
\end{bmatrix},\ \ \ \mbox{where $\dot{W} := dW(\epsilon)/d\epsilon |_{\epsilon=0}$.}
\end{align}
\end{lem}


Now we are ready to prove Theorem~\ref{thm:DetAlg}.  The necessity
argument follows from the one for \cite[Theorem~2]{CaiIsh:TAC10};
indeed, the class of strongly connected digraphs characterizes the
existence of a distributed algorithm that can solve average
consensus. For the sufficiency part, let
\begin{align} \label{eq:M_0_F}
M_0 :=\begin{bmatrix}
  I-L & 0 \\
  L & S
\end{bmatrix} \ \ \ \mbox{and} \ \ \ F :=\begin{bmatrix}
  0 & I \\
  0 & -I
\end{bmatrix}.
\end{align}
Then $M = M_0 + \epsilon F$, and we view $M$ as being obtained by
``perturbing'' $M_0$ via the term $\epsilon F$. Also, it is clear
that $M$ depends smoothly on $\epsilon$. Concretely, we will first
show that the eigenvalues $\lambda_i$ of the unperturbed matrix
$M_0$ satisfy
\begin{equation} \label{eq:eig_M0}
1 = \lambda_1 = \lambda_2 > |\lambda_3| \geq \cdots \geq
|\lambda_{2n}|.
\end{equation}
Then using Lemma~\ref{lem:eigpur} we will establish that after a
small perturbation $\epsilon F$, the obtained matrix $M$ has only a
simple eigenvalue $1$ and all other eigenvalues have moduli smaller
than one.  This is the characteristic part of our proof. Finally, it
follows from Proposition~\ref{prop:Mean_AveCon} that average
consensus is achieved. It should be pointed out that, unlike the
standard consensus algorithm (\ref{eq:stand_consen}), the tools in
nonnegative matrix theory cannot be used to analyze the spectrum of
$M$ directly due to the existence of negative entries.

\emph{Proof of Theorem~\ref{thm:DetAlg}.} (Necessity) Suppose that
$\mathcal {G}$ is not strongly connected. Then at least one node of
$\mathcal {G}$ is not globally reachable. Let $\mathcal {V}_g^*$
denote the set of non-globally reachable nodes, and write its
cardinality card$(\mathcal {V}_g^*) = r,\ r \in [1,n]$.  If $r = n$,
i.e. $\mathcal {G}$ does not have a globally reachable node, then
$\mathcal {G}$ has at least two distinct closed strong components
\cite[Theorem 2.1]{Lin:Mono:08}. In this case, if the nodes in
different components have different initial states, then average
consensus cannot be achieved.  It is left to consider $r<n$. Let
$\mathcal {V}_g := \mathcal {V} - \mathcal {V}_g^*$ denote the set
of all globally reachable nodes; thus $\mathcal {V}_g$ is the unique
closed strong component in $\mathcal {G}$ \cite[Theorem
2.1]{Lin:Mono:08}. Consider an initial condition $(x(0),0)$ such
that all nodes in $\mathcal {V}_g$ have the same state $c \in
\mathbb{R}$, and not all the states of the nodes in $\mathcal
{V}_g^*$ equal $c$. Hence $x_a \neq c$.  But no state or surplus
update is possible for the nodes in $\mathcal {V}_g$ because it is
closed, and therefore average consensus cannot be achieved.

(Sufficiency) First, we prove the assertion (\ref{eq:eig_M0}). Since
$M_0$ is block (lower) triangular, its spectrum is
$\sigma(M_0)=\sigma(I-L) \cup \sigma(S)$. Recall that the matrices
$I-L$ and $S$ are row and column stochastic, respectively; so their
spectral radii satisfy $\rho(I-L)=\rho(S)=1$. Now owing to that
$\mathcal {G}$ is strongly connected, $I-L$ and $S$ are both
\emph{irreducible}; thus by the Perron-Frobenius Theorem (see, e.g.,
\cite[Chapter 8]{HorJoh:90}) $\rho(I-L)$ (resp. $\rho(S)$) is a
simple eigenvalue of $I-L$ (resp. $S$). This implies
(\ref{eq:eig_M0}). Moreover, for $\lambda_1 = \lambda_2 = 1$, one
derives that the corresponding geometric multiplicity equals two by
verifying rank$(M_0-I)=2n-2$. Hence the eigenvalue $1$ is
semi-simple.

Next, we will qualify the changes of the semi-simple eigenvalue
$\lambda_1=\lambda_2=1$ of $M_0$ under a small perturbation
$\epsilon F$.  We do this by computing the derivatives
$d\lambda_1(\epsilon)/d\epsilon$ and
$d\lambda_2(\epsilon)/d\epsilon$ using Lemma~\ref{lem:eigpur}; here
$\lambda_1(\epsilon)$ and $\lambda_2(\epsilon)$ are the eigenvalues
of $M$ corresponding respectively to $\lambda_1$ and $\lambda_2$. To
that end, choose the right eigenvectors $y_1,y_2$ and left eigenvectors $z_1,z_2$ of the semi-simple
eigenvalue $1$ as follows:
\begin{align*} 
Y := \begin{bmatrix}
  y_1 & y_2
\end{bmatrix}= \begin{bmatrix}
  0 & \textbf{1} \\
  v_2 & -n v_2
\end{bmatrix},\ \ \
Z := \begin{bmatrix}
  z_1^T \\ z_2^T
\end{bmatrix}= \begin{bmatrix}
  \textbf{1}^T & \textbf{1}^T \\
  v_1^T & 0
\end{bmatrix}.
\end{align*}
Here $v_1 \in \mathbb{R}^n$ is a left eigenvector of $I-L$ with
respect to $\rho(I-L)$ such that it is positive and scaled to
satisfy $v_1^T \textbf{1}=1$; and $v_2 \in \mathbb{R}^n$ is a right
eigenvector of $S$ corresponding to $\rho(S)$ such that it is
positive and scaled to satisfy $\textbf{1}^T v_2=1$. The fact that
positive eigenvectors $v_1$ and $v_2$ exist follows again from the
Perron-Frobenius Theorem.  With this choice, one readily checks $Z Y
= I$. Now since $dM/d\epsilon |_{\epsilon=0} = F$, the matrix
(\ref{matrix:eigpur}) in the present case is
\begin{align*}
\begin{bmatrix}
  z_1^T F y_1 & z_1^T F y_2 \\
  z_2^T F y_1 & z_2^T F y_2
\end{bmatrix} = \begin{bmatrix}
  0 & 0 \\
  v_1^T v_2 & -n v_1^T v_2
\end{bmatrix}.
\end{align*}
It follows from Lemma~\ref{lem:eigpur} that for small $\epsilon>0$,
the derivatives $d\lambda_1(\epsilon)/d\epsilon$,
$d\lambda_2(\epsilon)/d\epsilon$ exist and are the eigenvalues of
the above matrix. Hence $d\lambda_1(\epsilon)/d\epsilon = 0$, and
$d\lambda_2(\epsilon)/d\epsilon = -n v_1^T v_2 <0$. This implies
that when $\epsilon$ is small, $\lambda_1(\epsilon)$ stays put while
$\lambda_2(\epsilon)$ moves to the left along the real axis.
Then by continuity, there must exist a positive $\delta_1$ such that
$\lambda_1(\delta_1)=1$ and $\lambda_2(\delta_1)<1$.  On the other
hand, since eigenvalues are continuous functions of matrix entries
(e.g., \cite{Bha:96,SteSun:90}), there must exist a positive
$\delta_2$ such that $|\lambda_i(\delta_2)|<1$ for all $i \in
[3,2n]$.  Thus for any sufficiently small $\epsilon \in
(0,\min\{\delta_1, \delta_2\})$, the matrix $M$ has a simple
eigenvalue $1$ and all other eigenvalues have moduli smaller than
one.  Therefore, from Proposition~\ref{prop:Mean_AveCon}, the
conclusion that average consensus is achieved follows. \hfill
$\blacksquare$

\begin{rem} \label{rem:confac_d}
Assuming that the deterministic algorithm (\ref{eq:DetAlg})
converges to the average, the speed of its convergence is governed
by the second largest (in modulus) eigenvalue of the matrix $M$.  We
denote this particular eigenvalue by $\lambda_2^{(d)}$, and refer to
it as the \emph{convergence factor} of algorithm (\ref{eq:DetAlg}).
Note that $\lambda_2^{(d)}$ depends not only on the graph topology
but also on the parameter $\epsilon$, and $\lambda_2^{(d)}<1$ is
equivalent to average consensus (by
Proposition~\ref{prop:Mean_AveCon}).
\end{rem}

\begin{rem} \label{rem:tech1}
Because of adding surpluses, the matrix $M$ in (\ref{eq:DetAlg}) is double
in size and is not nonnegative.  Hence standard nonnegative matrix
tools cannot be directly applied; this point was also discussed in
\cite{FraGiuSea:09}.  In \cite{LiuAndCaoMor:09} a system matrix
containing negative entries was analyzed, which depends however on symmetry
of network structures.  By contrast, we deal with general network topologies
and have demonstrated that certain matrix perturbation tools are useful in
proving convergence.
\end{rem}


\subsection{Proof of Proposition~\ref{prop:perturb_bound}} \label{pf2}


Some preliminaries will be presented first, based on which
Proposition~\ref{prop:perturb_bound} follows immediately. Henceforth
in this subsection, the digraph $\mathcal {G}$ is assumed to be
strongly connected. We begin by introducing a metric for the
distance between the spectrums of $M_0$ and $M$; here $M = M_0 +
\epsilon F$, with $M_0$ and $F$ in (\ref{eq:M_0_F}). Let
$\sigma(M_0):=\{\lambda_1,\ldots,\lambda_{2n}\}$ (where the
numbering is the same as that in (\ref{eq:eig_M0})) and
$\sigma(M):=\{\lambda_1(\epsilon),\ldots,\lambda_{2n}(\epsilon)\}$.
The \emph{optimal matching distance}
$d\left(\sigma(M_0),\sigma(M)\right)$ \cite{Bha:96,SteSun:90} is
defined by
\begin{align}\label{eq:optmat}
d\left(\sigma(M_0),\sigma(M)\right) := \min_{\pi} \ \max_{i \in
[1,2n]} \ |\lambda_i - \lambda_{\pi(i)}(\epsilon)|,
\end{align}
where $\pi$ is taken over all permutations of $\{1,\ldots,2n\}$.
Thus if we draw $2n$ identical circles centered respectively at
$\lambda_1,\ldots,\lambda_{2n}$, then
$d\left(\sigma(M_0),\sigma(M)\right)$ is the smallest radius such
that these circles include all
$\lambda_1(\epsilon),\ldots,\lambda_{2n}(\epsilon)$. Here is an
upper bound on the optimal matching distance
\cite[Theorem~VIII.1.5]{Bha:96}.

\begin{lem} \label{lem:optdis}
$d\left(\sigma(M_0),\sigma(M)\right) \leq 4 \ (||M_0||_\infty +
||M||_\infty)^{1-1/n} \ ||\epsilon F||_\infty^{1/n}$.
\end{lem}

Next, we are concerned with the eigenvalues
$\lambda_3(\epsilon),\ldots,\lambda_{2n}(\epsilon)$ of $M$, whose
corresponding unperturbed counterparts
$\lambda_3,\ldots,\lambda_{2n}$ of $M_0$ lie strictly inside the
unit circle (see the proof of Theorem~\ref{thm:DetAlg}).

\begin{lem} \label{lem:perturb_bound_1} If the parameter
$\epsilon \in (0,\bar{\epsilon}^{(d)})$ with $\bar{\epsilon}^{(d)}$
in (\ref{eq:perturb_bound}), then
$|\lambda_3(\epsilon)|,\ldots,|\lambda_{2n}(\epsilon)|<1$.
\end{lem}
\begin{prf}
Since $L=D-A$ and $S=(I-\tilde{D})+B$, one can compute
$||L||_\infty=2 \max_{i\in [1,n]} d_i<2$ and $||S||_\infty < n$.
Then $ ||M_0||_\infty \leq ||L||_\infty+||S||_\infty < 2 + n$ and
$||F||_\infty \leq 1$.  By Lemma~\ref{lem:optdis},
\begin{align*}
d\left(\sigma(M_0),\sigma(M)\right) &\leq 4 \ (2||M_0||_\infty +
\epsilon||F||_\infty)^{1-1/n} \ (\epsilon ||F||_\infty)^{1/n} \\
&< 4 \ (4+2n+\epsilon)^{1-1/n} \ \epsilon^{1/n} < 4 \ (4+2n+\epsilon) \ \epsilon^{1/n} < 1-|\lambda_3|.
\end{align*}
The last inequality is due to $\epsilon < \bar{\epsilon}^{(d)}$ in
(\ref{eq:perturb_bound}). Now recall from the proof of
Theorem~\ref{thm:DetAlg} that the unperturbed eigenvalues
$\lambda_3,\ldots,\lambda_{2n}$ of $M_0$ lie strictly inside the
unit circle; in particular, (\ref{eq:eig_M0}) holds. Therefore,
perturbing the eigenvalues $\lambda_3,\ldots,\lambda_{2n}$ by an
amount less than $\bar{\epsilon}$, the resulting eigenvalues
$\lambda_3(\epsilon),\ldots,\lambda_{2n}(\epsilon)$ will remain
inside the unit circle. \hfill $\blacksquare$
\end{prf}

It is left to consider the eigenvalues $\lambda_{1}(\epsilon)$ and
$\lambda_{2}(\epsilon)$ of $M$. Since every column sum of $M$ equals
one for an arbitrary $\epsilon$, we obtain that $1$ is always an
eigenvalue of $M$. Hence $\lambda_{1}(\epsilon)$ must be equal to
$1$ for any $\epsilon$.  On the other hand, for
$\lambda_2(\epsilon)$ the following is true.

\begin{lem} \label{lem:perturb_bound} If the parameter
$\epsilon \in (0,\bar{\epsilon}^{(d)})$ with $\bar{\epsilon}^{(d)}$
in (\ref{eq:perturb_bound}), then $|\lambda_2(\epsilon)|<1$.
\end{lem}
\begin{prf}
First recall from the proof of Theorem~\ref{thm:DetAlg} that
$\lambda_2=1$ and $d\lambda_2(\epsilon)/d\epsilon<0$; so for sufficiently small
$\epsilon>0$, it holds that $|\lambda_2(\epsilon)|<1$. Now suppose
that there exists $\delta \in (0,\bar{\epsilon}^{(d)})$ such that
$|\lambda_2(\delta)|\geq 1$. Owing to the continuity of eigenvalues,
it suffices to consider $|\lambda_2(\delta)| = 1$. There are three
such possibilities, for each of which we derive a contradiction.

Case 1: $\lambda_2(\delta)$ is a complex number with nonzero
imaginary part and $|\lambda_2(\delta)| = 1$.  Since $M$ is a real
matrix, there must exist another eigenvalue $\lambda_i(\delta)$, for
some $i \in [3,2n]$, such that $\lambda_i(\delta)$ is a complex
conjugate of $\lambda_2(\delta)$. Then
$|\lambda_i(\delta)|=|\lambda_2(\delta)| = 1$, which is in
contradiction to that all the eigenvalues
$\lambda_3(\delta),\ldots,\lambda_{2n}(\delta)$ stay inside the unit
circle as $\delta \in (0,\bar{\epsilon}^{(d)})$ by
Lemma~\ref{lem:perturb_bound_1}.

Case 2: $\lambda_2(\delta)=-1$.  This implies at least
$d\left(\sigma(M_0),\sigma(M)\right)=2$, which contradicts
$d\left(\sigma(M_0),\sigma(M)\right) < 1-|\lambda_3|<1$ when
(\ref{eq:perturb_bound}) holds.

Case 3: $\lambda_2(\delta)=1$. This case is impossible because the
eigenvalue $1$ of $M$ is always simple, as we have justified in the
necessity proof of Proposition~\ref{prop:Mean_AveCon}. \hfill
$\blacksquare$

\end{prf}

Summarizing Lemmas~\ref{lem:perturb_bound_1} and
\ref{lem:perturb_bound}, we obtain that if the parameter $\epsilon
\in (0,\bar{\epsilon}^{(d)})$ with $\bar{\epsilon}^{(d)}$ in
(\ref{eq:perturb_bound}), then $\lambda_{1}(\epsilon)=1$ and
$|\lambda_2(\epsilon)|$, $|\lambda_3(\epsilon)|$, $\ldots$,
$|\lambda_{2n}(\epsilon)|<1$. Therefore, by
Proposition~\ref{prop:Mean_AveCon} the deterministic algorithm
(\ref{eq:DetAlg}) achieves average consensus; this establishes
Proposition~\ref{prop:perturb_bound}.

\section{Averaging in Asynchronous Networks} \label{Sec4_GosAlg}

We move on to solve Problem 2.  First, a surplus-based gossip
algorithm is designed for digraphs, which extends those algorithms
\cite{BoGhPrSh:06,KaBaSr:07,CaFaFrZa:Automatica10,LavMur:TAC10} only
for undirected graphs.  Then, mean-square and almost sure
convergence to average consensus is justified for arbitrary
strongly connected topologies.

\subsection{Algorithm Description}

Consider again a network of $n$ agents modeled by a digraph
$\mathcal {G} = (\mathcal {V}, \mathcal {E})$.  Suppose that at each
time, exactly one edge in $ \mathcal {E}$ is activated at random,
independently from all earlier instants. Say edge $(j,i)$ is
activated at time $k \in \mathbb{Z}_+$, with a constant probability
$p_{ij} \in (0,1)$. Along the edge, the state information $x_j(k)$
and surplus $s_j(k)$ are transmitted from node $j$ to $i$. The
induced update is described as follows:

\noindent (i) Let $w_{ij}\in (0,1)$ be the updating weight, and $\epsilon >0$ be a parameter.
For node $i$:
\begin{align} \label{eq:i_sta_update}
x_i(k+1) = x_i(k) + w_{ij}(x_j(k)-x_i(k)) + \epsilon w_{ij} s_i(k),
\end{align}
\begin{equation} \label{eq:i_sur_update}
\begin{split}
s_i(k+1) = s_i(k)+s_j(k) - (x_i(k+1)-x_i(k)),
\end{split}
\end{equation}

\noindent (ii) For node $j$: $x_j(k+1) = x_j(k)$ and $s_j(k+1) = 0$.

\noindent (iii) For other nodes $l \in \mathcal {V}-\{i,j\}$:
$x_l(k+1) = x_l(k)$ and $s_l(k+1) = s_l(k)$.

We discuss potential applications of this protocol in sensor
networks. Our focus is again on the situations of directed
information flow, like asynchronous communication with variable
ranges or unidirectional geographic routing
\cite{KemDobGeh:03,BBTTV:10}.  First, the states and surpluses can
be implemented as ordinary variables in sensors, since their
exchange and updating rules are fairly simple and purely local.
Also, we will see that the parameter $\epsilon$, as in the
algorithm~(\ref{eq:DetAlg}), affects the convergence of the
algorithm, and must be chosen to be sufficiently small.  A valid
upper bound for $\epsilon$ involves again non-local information of
the network; thus computing a bound offline and then notifying that
value to every node is one possible way to deal with this
restriction.


Now let $A_{ji}$ be the adjacency matrix of the digraph $\mathcal
{G}_{ji}=(\mathcal {V},\{(j,i)\})$ given by $A_{ji}=w_{ij}f_if_j^T$,
where $f_i$, $f_j$ are unit vectors of the standard basis of
$\mathbb{R}^n$.  Then the Laplacian matrix $L_{ji}$ is given by
$L_{ji}:=D_{ji}-A_{ji}$, where $D_{ji}=w_{ij}f_if_i^T$. Thus
$L_{ji}$ has zero row sums, and the matrix $I-L_{ji}$ is row
stochastic. Also define $S_{ji}:=I-(f_j-f_i)f_j^T$; it is clear that
$S_{ji}$ is column stochastic.  With these matrices, the
iteration of states and surpluses when edge $(j,i)$ is activated at
time $k$ can be written in the matrix form as
\begin{align} \label{eqa:alg_1}
\begin{bmatrix}
  x(k+1)\\
  s(k+1)
\end{bmatrix}=
M(k)
\begin{bmatrix}
  x(k)\\
  s(k)
\end{bmatrix},
\ \mbox{where } M(k)=M_{ji}:=
\begin{bmatrix}
  I-L_{ji} & \epsilon D_{ji}\\
  L_{ji} & S_{ji}-\epsilon D_{ji}
\end{bmatrix} .
\end{align}
We have several remarks regarding this algorithm.  (i) The matrix
$M(k)$ has negative entries due to the presence of the Laplacian
matrix $L_{ji}$ in the $(2,1)$-block. (ii) The column sums of $M(k)$
are equal to one, which implies that the quantity $x(k)+s(k)$ is
constant for all $k$.  (iii) By the assumption on the probability
distribution of activating edges, the sequence $M(k)$,
$k=0,1,\ldots$, is independent and identically distributed (i.i.d.).
Henceforth we refer to (\ref{eqa:alg_1}) as the gossip algorithm,
and establish its mean-square and almost sure convergence in the
sequel.

\begin{exmp} \label{ex:gosalg}
Consider again the network of four nodes in
Fig.~\ref{fig:Ex_4nodes}. We give one instance of the matrix $M(k)$
when edges $(3,2)$ is activated, with the updating weight $w_{23}=1/2$.
{\small
\begin{align*} M_{32}&=\left[
\begin{array}{cccc|cccc}
1 & 0 & 0 & 0 & 0 & 0 & 0 & 0 \\
0 & 1/2 & 1/2 & 0 & 0 & \epsilon/2 & 0 & 0 \\
0 & 0 & 1 & 0 & 0 & 0 & 0 & 0 \\
0 & 0 & 0 & 1 & 0 & 0 & 0 & 0 \\
\hline  0 & 0 & 0 & 0 & 1 & 0 & 0 & 0 \\
0 & 1/2 & -1/2 & 0 & 0 & 1- \epsilon/2 & 1 & 0 \\
0 & 0 & 0 & 0 & 0 & 0 & 0 & 0 \\
0 & 0 & 0 & 0 & 0 & 0 & 0 & 1
\end{array}\right],
\end{align*}
}
We see that $M(k)$ has negative entries, and every
column sums up to one.
\end{exmp}


\subsection{Convergence Result}

We present our main result in this section.

\begin{thm} \label{them:msqr}
Using the gossip algorithm (\ref{eqa:alg_1}) with the parameter
$\epsilon>0$ sufficiently small, the agents achieve mean-square
average consensus if and only if the digraph $\mathcal {G}$ is
strongly connected.
\end{thm}

We remark that Theorem~\ref{them:msqr} generalizes the convergence
result in \cite{BoGhPrSh:06} from undirected to directed graphs. The
problem of achieving average consensus on gossip digraphs is,
however, more difficult in that the state sum of the nodes need not
be invariant at each iteration.  The key in our extension is to
augment surplus variables which keep track of individual state
updates, thereby ensuring average consensus for general strongly
connected digraphs.  This approach was previously exploited in \cite{FraGiuSea:09}
for a broadcast gossip algorithm, however without a convergence proof.
We remark that our technique to prove Theorem~\ref{them:msqr}, based on matrix perturbation theory,
can be applied to \cite{FraGiuSea:09} and justify the algorithm convergence.

We note that in the literature, many works for agents with non-scalar
dynamics deal only with static networks (e.g., \cite{LDCH:10,CaoRen:10}).
Some exceptions include \cite{LiuAndCaoMor:09} which relies heavily on
graph symmetry and \cite{RenBea:08} which is based on dwell-time switching.
By contrast, we study general digraphs that switch at every discrete
time and each resulting update matrix is not nonnegative. The corresponding
analysis is difficult, and we will demonstrate again that matrix perturbation
tools are instrumental in proving convergence.



To prove Theorem~\ref{them:msqr}, three preliminary results are to
be established in order. The first is a necessary and sufficient
condition for mean-square average consensus characterized by the
spectrum of the matrix $E\left[ M(k) \otimes M(k) \right]$, where
$\otimes$ stands for the Kronecker product.  This condition will be
used in the sufficiency proof of Theorem~\ref{them:msqr}. Since the
matrices $M(k)$ are i.i.d. we denote $E\left[ M(k) \otimes M(k)
\right]$ by $E\left[ M \otimes M \right]$.  This result corresponds
to Proposition~\ref{prop:Mean_AveCon} for the deterministic
algorithm in Section~\ref{Sec3_SynAlg}.


\begin{prop} \label{prop:msqr}
The gossip algorithm (\ref{eqa:alg_1}) achieves mean-square average
consensus if and only if $1$ is a simple eigenvalue of $E\left[ M
\otimes M \right]$, and all the other eigenvalues have moduli
smaller than one.
\end{prop}
\begin{prf}
(Sufficiency)
Define the consensus error $e(k)$, $k \geq 0$, as
\begin{align} \label{eq:con_err}
e(k) := \begin{bmatrix}
  x(k)\\
  s(k)
\end{bmatrix}-\begin{bmatrix}
  x_a \textbf{1}\\
  0
\end{bmatrix} \in \mathbb{R}^{2n}.
\end{align}
We must show that $E\left[ e(k)^T e(k) \right] \rightarrow 0$ as $k
\rightarrow \infty$.  Since $\textbf{1}^T (x(k)+s(k)) = \textbf{1}^T
x(0)$ for every $k\geq 0$, $e(k)$ is orthogonal to $[\textbf{1}^T \
\textbf{1}^T]^T$ (i.e., $[\textbf{1}^T \ \textbf{1}^T] e(k)=0$).
Also it is easy to check $e(k+1)=M(k)e(k)$; thus
$e(k+1)e(k+1)^T = M(k) e(k)e(k)^T M(k)^T.$
Collect the entries of $e(k)e(k)^T$, drawn column wise, into a
vector $\tilde{e}(k) \in \mathbb{R}^{4n^2}$.  It then suffices to
show that $E\left[ \tilde{e}(k) \right] \rightarrow 0$ as $k
\rightarrow \infty$.

Now it follows that
$\tilde{e}(k+1)= \left( M(k) \otimes M(k) \right) \tilde{e}(k)$ (cf.
\cite{BoGhPrSh:06}).
Hence $E\left[ \tilde{e}(k+1) | \tilde{e}(k) \right] = E\left[ M
\otimes M \right] \tilde{e}(k)$, and condition repeatedly to obtain
$E\left[ \tilde{e}(k) \right] = E\left[ M \otimes M \right]^k
\tilde{e}(0)$.  Note that every column of $E\left[ M \otimes M
\right]$ sums up to one, and
\begin{align*} E\left[ M \otimes M \right]
\left( \begin{bmatrix}
  \textbf{1}\\
  0
\end{bmatrix} \otimes \begin{bmatrix}
  \textbf{1}\\
  0
\end{bmatrix} \right) =
\begin{bmatrix}
  \textbf{1}\\
  0
\end{bmatrix} \otimes \begin{bmatrix}
  \textbf{1}\\
  0
\end{bmatrix};
\end{align*}
so $1$ is an eigenvalue of $E\left[ M \otimes M \right]$, with
$[\textbf{1}^T \ \textbf{1}^T]^T \otimes [\textbf{1}^T \
\textbf{1}^T]^T$ and $[\textbf{1}^T \ 0]^T \otimes [\textbf{1}^T \
0]^T$ as associated left and right eigenvectors, respectively. Write
$E\left[ M \otimes M \right]$ in Jordan canonical form as
\begin{align*}
E\left[ M \otimes M \right] = V J V^{-1} =\begin{bmatrix}
  y_1 & \cdots & y_{4n^2}
\end{bmatrix}
\begin{bmatrix}
  1 & 0 \\
  0 & J'
\end{bmatrix}
\begin{bmatrix}
  z_1^T \\ \vdots \\ z_{4n^2}^T
\end{bmatrix},
\end{align*}
where $J'$ contains the Jordan block matrices corresponding to those
eigenvalues with moduli smaller than one. For the eigenvalue $1$
choose $y_1=[\textbf{1}^T \ 0]^T \otimes [\textbf{1}^T \ 0]^T$ and
$z_1=1/n^2[\textbf{1}^T \ \textbf{1}^T]^T \otimes [\textbf{1}^T \
\textbf{1}^T]^T$; thus $z_1^T y_1 = 1$.
Then the $k$th power of $E\left[ M \otimes M \right]$ is
\begin{align*}
E\left[ M \otimes M \right]^k = V J^k V^{-1}
    = V
\begin{bmatrix}
  1 & 0 \\
  0 & (J')^k
\end{bmatrix} V^{-1} \rightarrow y_1 z_1^T,\ \ \
\mbox{as $k \rightarrow \infty$.}
\end{align*}
Therefore we obtain
\begin{align*}
E\left[ \tilde{e}(k) \right] \rightarrow y_1 z_1^T \tilde{e}(0)  =
y_1 \sum_{i=1}^{2n} \left( e_i(0) \sum_{j=1}^{2n} e_j(0) \right) =
y_1 \sum_{i=1}^{2n} e_i(0) \cdot 0 = 0,
\end{align*}
where the second equality is due to $e(k)\ \bot\ [\textbf{1}^T \
\textbf{1}^T]^T$.

(Necessity)  Suppose $E\left[ e(k)^T e(k) \right] \rightarrow 0$ as
$k \rightarrow \infty$.  Then $E\left[ e_i(k)^2 \right] \rightarrow
0$ for all $i$. It thus follows from the Cauchy-Schwartz inequality
(e.g., \cite{GriSti:01}) that $E\left[ |e_i(k) e_j(k)| \right]^2
\leq E\left[ e_i(k)^2 \right] E\left[ e_j(k)^2 \right] \rightarrow
0$, for every $i,j\in [1,2n]$. This implies $E\left[ \tilde{e}(k)
\right] \rightarrow 0$; so $\lim_{k \rightarrow \infty} E\left[ M
\otimes M \right]^k \tilde{e}(0) = 0$. Also, it is known
\cite{XiBo:04} that $\lim_{k \rightarrow \infty} E\left[ M \otimes M
\right]^k$ exists if and only if there is a nonsingular $V$ such
that
\begin{align*}
E\left[ M \otimes M \right] = V J V^{-1} =\begin{bmatrix}
  y_1 & \cdots & y_{4n^2}
\end{bmatrix}
\begin{bmatrix}
  I_\kappa & 0 \\
  0 & J'
\end{bmatrix}
\begin{bmatrix}
  z_1^T \\ \vdots \\ z_{4n^2}^T
\end{bmatrix},
\end{align*}
where $\kappa \in [1,2n]$ and $\rho(J')<1$. Hence $\lim_{k
\rightarrow \infty} E\left[ M \otimes M \right]^k \tilde{e}(0)=
\left( \sum_{i=1}^\kappa y_i z_i^T \right)\tilde{e}(0) = 0$. Now
suppose $\kappa>1$. Choose as before $z_1=1/n^2[\textbf{1}^T \
\textbf{1}^T]^T \otimes [\textbf{1}^T \ \textbf{1}^T]^T$, and recall
$z_1^T e(0)=0$. We know from the structure of $J$ that for every
$j\in[2,\kappa]$, $z_j$ is linearly independent of $z_1$, which
indicates $z_j^T e(0) \neq 0$ and consequently $\left(
\sum_{i=1}^\kappa y_i z_i^T \right)\tilde{e}(0) \neq 0$.  Therefore
$\kappa=1$, i.e., the eigenvalue $1$ of $E\left[ M \otimes M
\right]$ is simple and all the others have moduli smaller than one.
\hfill $\blacksquare$
\end{prf}


The second preliminary is an easy corollary of the Perron-Frobenius
Theorem.

\begin{lem}(cf. \cite[Chapter~XIII]{Gan:59}) \label{lem:pern_frob_cor}
Let $W$ be a nonnegative and irreducible matrix, and $\lambda$ be an
eigenvalue of $W$.  If there is a positive vector $v$ such that $W
v=\lambda v$, then $\lambda=\rho(W)$.
\end{lem}
\begin{prf}
Since $W$ is nonnegative and irreducible, the Perron-Frobenius
Theorem implies that $\rho(W)$ is a simple eigenvalue of $W$ and
there is a positive left eigenvector $w$ corresponding to $\rho(W)$,
i.e., $w^T W=w^T \rho(W)$.  Then
\begin{align*}
\rho(W) (v^T w) = v^T (\rho(W)w) = v^T (W^T w) = (Wv)^T w = (\lambda
v)^T w = \lambda (v^T w),
\end{align*}
which yields $(\lambda-\rho(W)) (v^T w)=0$.  Since both $v$ and $w$
are positive, we conclude that $\lambda=\rho(W)$. \hfill
$\blacksquare$
\end{prf}


The last preliminary is on the spectral properties of the following
four matrices: $E\left[ (I-L) \otimes (I-L) \right]$, $E\left[ (I-L)
\otimes S \right]$, $E\left[ S \otimes (I-L) \right]$, and $E\left[
S \otimes S \right]$.

\begin{lem} \label{lem:msqr} Suppose
that the digraph $\mathcal {G}=(\mathcal {V},\mathcal {E})$ is
strongly connected. Then each of the four matrices $E\left[ (I-L)
\otimes (I-L) \right]$, $E\left[ (I-L) \otimes S \right]$, $E\left[
S \otimes (I-L) \right]$, and $E\left[ S \otimes S \right]$ has a
simple eigenvalue $1$ and all other eigenvalues with moduli smaller
than one.
\end{lem}
\noindent \emph{Proof.} First observe that all the four matrices are
nonnegative, for $I-L_{ji}$ and $S_{ji}$ are for every $(j,i)\in
\mathcal {E}$.  Then since $(I-L_{ji})\textbf{1}=\textbf{1}$ and
$\textbf{1}^TS_{ji}=\textbf{1}^T$ for every $(j,i)\in \mathcal {E}$,
a short calculation yields the following:
\begin{align*}
& E\left[ (I-L) \otimes (I-L) \right] (\textbf{1} \otimes
\textbf{1}) = (\textbf{1} \otimes \textbf{1});\ \ \ \ \ E\left[ (I-L) \otimes S \right] (\textbf{1} \otimes
v_2) = (\textbf{1} \otimes v_2);\\
& (\textbf{1}^T \otimes v_1^T) E\left[ S \otimes (I-L) \right]  =
(\textbf{1}^T \otimes v_1^T);\ \ \ \ \ (\textbf{1}^T \otimes \textbf{1}^T) E\left[ S \otimes S \right]  =
(\textbf{1}^T \otimes \textbf{1}^T).
\end{align*}
Here $v_1$ is positive such that $v_1^T E\left[ I-L \right]=v_1^T$
and $v_1^T \textbf{1}=1$, and $v_2$ is positive such that $E\left[ S
\right]v_2=v_2$ and $\textbf{1}^Tv_2=1$. Thus each matrix has an
eigenvalue $1$, and the corresponding right or left eigenvector is
positive.  In what follows, it will be shown that all the four
matrices are irreducible.  Then the conclusion will follow from
Lemma~\ref{lem:pern_frob_cor} and the Perron-Frobenius Theorem.

We first prove that $E\left[ (I-L) \otimes (I-L) \right]$ is
irreducible, which is equivalent to that the digraph $\hat{\mathcal
{G}}=(\hat{\mathcal {V}},\hat{\mathcal {E}})$ corresponding to this
matrix is strongly connected, where $\hat{\mathcal {V}}:=\mathcal
{V} \times \mathcal {V} =\{(i,i') : i,i' \in \mathcal {V}\}$.
Arrange the nodes in $\hat{\mathcal {V}}$ so that $\hat{\mathcal
{V}}=\mathcal {V}_1 \cup \cdots \cup \mathcal {V}_n$, where
$\mathcal {V}_p=\{(p,1),\ldots,(p,n)\}$ for every $p\in[1,n]$.  Now
since $E\left[ (I-L) \otimes (I-L) \right] = \sum_{(j,i)\in\mathcal
{E}} p_{ij} (I-L_{ji}) \otimes (I-L_{ji})$, the digraph
$\hat{\mathcal {G}}$ is the union of the digraphs corresponding to
$p_{ij} (I-L_{ji}) \otimes (I-L_{ji})$. Note that each $p_{ij}
(I-L_{ji}) \otimes (I-L_{ji})$ gives rise to (i) an edge from
$(p,j)$ to $(p,i)$ in $\mathcal {V}_p$ for every $p\in[1,n]$, and
(ii) edges from some nodes in $\mathcal {V}_j$ to some nodes in
$\mathcal {V}_i$.  Owing to that $\mathcal {G}$ is strongly
connected, the union of the above edges yields, for every $i,j \in
[1,n]$, (i) a directed path from $(p,i)$ to $(p,j)$ in $\mathcal
{V}_p$ for every $p\in[1,n]$, and (ii) directed paths from some
nodes in $\mathcal {V}_i$ to some nodes in $\mathcal {V}_j$. This
implies that there is a directed path from $(p,i)$ to $(q,j)$ for
every $p,q,i,j\in[1,n]$, i.e., $\hat{\mathcal {G}}$ is strongly
connected, and hence $E\left[ (I-L) \otimes (I-L) \right]$ is
irreducible.

By a similar argument, we derive that the digraphs corresponding to
$E\left[ (I-L) \otimes S \right]$, $E\left[ S \otimes (I-L)
\right]$, and $E\left[ S \otimes S \right]$ are all strongly
connected.  Therefore they are also irreducible. \hfill
$\blacksquare$


We are now ready to provide the proof of Theorem~\ref{them:msqr}.
The necessity argument is the same as Theorem~\ref{thm:DetAlg}.
Below is the sufficiency part.

\emph{Proof of Theorem~\ref{them:msqr}.} (Sufficiency) By
Proposition~\ref{prop:msqr} it suffices to show that the matrix
$E\left[ M \otimes M \right]$ has a simple eigenvalue $1$, and all
other eigenvalues with moduli smaller than one.  Let $M_0(k)
:=\begin{bmatrix}
  I-L(k) & 0 \\
  L(k) & S(k)
\end{bmatrix}$ and $F(k) :=\begin{bmatrix}
  0 & D(k) \\
  0 & -D(k)
\end{bmatrix}$; from (\ref{eqa:alg_1}) we have $M(k)=M_0(k)+\epsilon F(k)$.
Then write
\begin{align*} E\left[ M \otimes M \right] &= E\left[ (M_0+\epsilon
F) \otimes (M_0+\epsilon F) \right] = E\left[ M_0 \otimes M_0
\right] + \epsilon E\left[ M_0 \otimes F + F \otimes M_0 + F \otimes
\epsilon F\right]\\
&=E\Bigg\{ \begin{bmatrix}
  I-L & 0 \\
  L & S
\end{bmatrix} \otimes \begin{bmatrix}
  I-L & 0 \\
  L & S
\end{bmatrix} \Bigg\} + \epsilon E\Bigg\{ \begin{bmatrix}
  I-L & 0 \\
  L & S
\end{bmatrix} \otimes \begin{bmatrix}
  0 & D \\
  0 & -D
\end{bmatrix} +\\  &\mbox{\hspace{1.05cm}}\begin{bmatrix}
  0 & D \\
  0 & -D
\end{bmatrix} \otimes \begin{bmatrix}
  I-L & 0 \\
  L & S
\end{bmatrix} + \begin{bmatrix}
  0 & D \\
  0 & -D
\end{bmatrix} \otimes \epsilon \begin{bmatrix}
  0 & D \\
  0 & -D
\end{bmatrix} \Bigg\}.
\end{align*}
Let $p \in [1,4n]$, and $p\textbf{n} := \{(p-1)n+1,\ldots,pn\}$.
Consider the following permutation: 
\begin{align*} \{ & \textbf{n}, 3\textbf{n},
\ldots, (2n-1)\textbf{n};\ 2\textbf{n}, 4\textbf{n},
\ldots, 2n\textbf{n}; \\
&(2n+1)\textbf{n}, (2n+3)\textbf{n}, \dots, (4n-1)\textbf{n};\
(2n+2)\textbf{n}, (2n+4)\textbf{n}, \ldots, 4n\textbf{n} \}.
\end{align*}
Denoting by $P$ the corresponding permutation matrix (which is
orthogonal), one derives that
\begin{align}\label{eq:M0eF} P^T E\left[ M \otimes M \right] P &= P^T E\left[ M_0 \otimes M_0
\right] P + \epsilon P^T E\left[ M_0 \otimes F + F \otimes M_0 + F
\otimes \epsilon F\right] P =: \hat{M_0} + \epsilon \hat{F},
\end{align}
where
\begin{align*}
&\hat{M_0} := E\begin{bmatrix}
  (I-L) \otimes (I-L) & 0 & 0 & 0 \\
  (I-L) \otimes L & (I-L) \otimes S & 0 & 0 \\
  L \otimes (I-L) & 0 & S \otimes (I-L) & 0 \\
  L \otimes L & L \otimes S & S \otimes L & S \otimes S
\end{bmatrix},\\
&\hat{F} := E\begin{bmatrix}
  0 &  (I-L) \otimes D & D \otimes (I-L) & D \otimes \epsilon D \\
  0 & -(I-L) \otimes D & D \otimes  L    & D \otimes (S - \epsilon D) \\
  0 &     L  \otimes D &-D \otimes (I-L) & (S - \epsilon D) \otimes D \\
  0 &    -L  \otimes D &-D \otimes  L    & D \otimes
  (\epsilon D - S) - S \otimes D
\end{bmatrix}.
\end{align*}

Based on the above similarity transformation, we henceforth analyze
the spectral properties of the matrix $\hat{M_0} + \epsilon
\hat{F}$. For this, we resort again to a perturbation argument,
which proceeds similarly to the one for Theorem~\ref{thm:DetAlg}.
First, since the digraph $\mathcal {G}$ is strongly connected, it
follows from Lemma~\ref{lem:msqr} that the eigenvalues of the matrix
$\hat{M_0}$ satisfy
\begin{align} \label{eq:lambda5}
1 = \hat{\lambda}_1 = \hat{\lambda}_2 = \hat{\lambda}_3 =
\hat{\lambda}_4
> |\hat{\lambda}_5| \geq \cdots \geq |\hat{\lambda}_{4n^2}|.
\end{align}
For the eigenvalue $1$, one derives that the corresponding geometric
multiplicity equals four by verifying rank$(\hat{M_0}-I)=4n^2-4$.
Thus $1$ is a semi-simple eigenvalue.

Next, we will qualify the changes of the semi-simple eigenvalue
$\hat{\lambda}_1 = \hat{\lambda}_2 = \hat{\lambda}_3 =
\hat{\lambda}_4=1$ of $\hat{M_0}$ under a small perturbation
$\epsilon \hat{F}$. We do this by computing the derivatives
$d\hat{\lambda}_i(\epsilon)/d\epsilon$, $i\in [1,4]$, using
Lemma~\ref{lem:eigpur}; here $\hat{\lambda}_i(\epsilon)$ are the
eigenvalues of $\hat{M_0} + \epsilon \hat{F}$ corresponding to
$\hat{\lambda}_i$. To that end, choose the right and left
eigenvectors of the semi-simple eigenvalue $1$ as follows:
\begin{align*}
&Y := \begin{bmatrix}
  y_1 & y_2 & y_3 & y_4
\end{bmatrix}= \begin{bmatrix}
  0 & 0 & 0 & \textbf{1}\otimes\textbf{1} \\
  0 & 0 & \textbf{1}\otimes nv_2 & -\textbf{1}\otimes nv_2 \\
  0 & nv_2\otimes\textbf{1} & 0 & -nv_2\otimes\textbf{1} \\
  nv_2\otimes nv_2 & -nv_2\otimes nv_2 & -nv_2\otimes nv_2 & nv_2\otimes nv_2
\end{bmatrix}, \\
&Z := \begin{bmatrix}
  z_1^T \\ z_2^T \\ z_3^T \\ z_4^T
\end{bmatrix}= \begin{bmatrix}
  \frac{1}{n}\textbf{1}^T \otimes \frac{1}{n}\textbf{1}^T & \frac{1}{n}\textbf{1}^T \otimes\frac{1}{n}\textbf{1}^T & \frac{1}{n}\textbf{1}^T \otimes\frac{1}{n}\textbf{1}^T & \frac{1}{n}\textbf{1}^T \otimes\frac{1}{n}\textbf{1}^T\\
  \frac{1}{n}\textbf{1}^T \otimes v_1^T & 0 & \frac{1}{n}\textbf{1}^T \otimes v_1^T & 0\\
  v_1^T\otimes\frac{1}{n}\textbf{1}^T & v_1^T\otimes\frac{1}{n}\textbf{1}^T & 0 & 0 \\
  v_1^T\otimes v_1^T & 0 & 0 & 0
\end{bmatrix}.
\end{align*}
Here $v_1$ is positive such that $v_1^T E\left[ I-L \right]=v_1^T$
and $v_1^T \textbf{1}=1$, and $v_2$ is positive such that $E\left[ S
\right]v_2=v_2$ and $\textbf{1}^Tv_2=1$. With this choice, it is
readily checked that $Z Y = I$. Now the matrix $\hat{M_0} + \epsilon
\hat{F}$ depends smoothly on $\epsilon$, and the derivative
$d(\hat{M_0} + \epsilon \hat{F})/d\epsilon |_{\epsilon=0}$ is
\begin{align*}
\hat{F}_0 := \frac{d(\hat{M_0} + \epsilon \hat{F})}{d\epsilon}
\Bigg|_{\epsilon=0} &=
\left( \hat{F}+ \epsilon \frac{d \hat{F}}{d\epsilon} \right) \Bigg|_{\epsilon=0} \\
&= E\begin{bmatrix}
  0 &  (I-L) \otimes D & D \otimes (I-L) & 0 \\
  0 & -(I-L) \otimes D & D \otimes  L    & D \otimes S  \\
  0 &     L  \otimes D &-D \otimes (I-L) & S \otimes D \\
  0 &    -L  \otimes D &-D \otimes  L    & - D \otimes S - S \otimes D
\end{bmatrix}.
\end{align*}
Hence the matrix (\ref{matrix:eigpur}) in the present case is
\begin{align*}
\begin{bmatrix}
  z_1^T \hat{F}_0 y_1 & z_1^T \hat{F}_0 y_2 & z_1^T \hat{F}_0 y_3 & z_1^T \hat{F}_0 y_4 \\
  z_2^T \hat{F}_0 y_1 & z_2^T \hat{F}_0 y_2 & z_2^T \hat{F}_0 y_3 & z_2^T \hat{F}_0 y_4 \\
  z_3^T \hat{F}_0 y_1 & z_3^T \hat{F}_0 y_2 & z_3^T \hat{F}_0 y_3 & z_3^T \hat{F}_0 y_4 \\
  z_4^T \hat{F}_0 y_1 & z_4^T \hat{F}_0 y_2 & z_4^T \hat{F}_0 y_3 & z_4^T \hat{F}_0 y_4
\end{bmatrix} = \begin{bmatrix}
  0 & 0 & 0 & 0 \\
  n v_1^T E[D] v_2 & -n v_1^T E[D] v_2 & 0 & 0 \\
  n v_1^T E[D] v_2 & 0 & -n v_1^T E[D] v_2 & 0\\
  0 & n v_1^T E[D] v_2 & n v_1^T E[D] v_2 & -2 n v_1^T E[D] v_2
\end{bmatrix}.
\end{align*}
It follows from Lemma~\ref{lem:eigpur} that for small $\epsilon>0$,
the derivatives $d\hat{\lambda}_i(\epsilon)/d\epsilon$, $i\in
[1,4]$, exist and are the eigenvalues of the above matrix. Hence
$d\hat{\lambda}_1(\epsilon)/d\epsilon = 0$,
$d\hat{\lambda}_2(\epsilon)/d\epsilon =
d\hat{\lambda}_3(\epsilon)/d\epsilon = -n v_1^T E[D] v_2 <0$, and
$d\hat{\lambda}_4(\epsilon)/d\epsilon = -2 n v_1^T E[D] v_2 < 0$.
This implies that when $\epsilon$ is small,
$\hat{\lambda}_1(\epsilon)$ stays put, while
$\hat{\lambda}_2(\epsilon)$, $\hat{\lambda}_3(\epsilon)$, and
$\hat{\lambda}_4(\epsilon)$ move to the left along the real axis. So
by continuity, there exists a positive $\delta_1$ such that
$\lambda_1(\delta_1)=1$ and $\lambda_2(\delta_1)$,
$\lambda_3(\delta_1)$, $\lambda_4(\delta_1)<1$.  On the other hand,
by the eigenvalue continuity there exists a positive $\delta_2$ such
that $|\lambda_i(\delta_2)|<1$ for all $i \in [5,4n^2]$.  Therefore
for any sufficiently small $\epsilon \in (0,\min\{\delta_1,
\delta_2\})$, the matrix $\hat{M_0} + \epsilon \hat{F}$ has a simple
eigenvalue $1$ and all other eigenvalues with moduli smaller than
one. \hfill $\blacksquare$

\begin{rem} \label{rem:confac_g}
Assuming that the gossip algorithm (\ref{eqa:alg_1}) converges to
the average in mean square, the speed of its convergence is
determined by the second largest (in modulus) eigenvalue of the
matrix $E\left[ M \otimes M \right]$.  We denote this particular
eigenvalue by $\lambda_2^{(g)}$, and refer to it as the
\emph{convergence factor} of algorithm (\ref{eqa:alg_1}).  Notice
that $\lambda_2^{(g)}$ depends not only on the graph topology but
also on the parameter $\epsilon$, and $\lambda_2^{(g)}<1$ is
equivalent to mean-square average consensus (by
Proposition~\ref{prop:msqr}).
\end{rem}

\begin{rem} \label{rem:bnd_gos}
We have established that for small enough $\epsilon$, the gossip algorithm~
(\ref{eqa:alg_1}) achieves mean-square average consensus. Using the same
notion of optimal matching distance and following the procedures as in Subsection~\ref{pf2},
it may be possible to derive a general bound for $\epsilon$ by solving the
inequality $4 \ (||\hat{M}_0||_\infty + ||\hat{M}_0+\epsilon \hat{F}||_\infty)^{1-1/n} \ ||\epsilon \hat{F}||_\infty^{1/n} < 1 - |\hat{\lambda}_5|$, where $\hat{M}_0, \hat{F}$ are from (\ref{eq:M0eF}) and $\hat{\lambda}_5$ from (\ref{eq:lambda5}).
The corresponding computation is however rather long, since the
involved matrices are of much larger sizes.
Such a general bound unavoidably again involves $n$, the number of agents
in the network, and $\hat{\lambda}_5$, the second
largest eigenvalue of one of the four matrices in Lemma~\ref{lem:msqr}.
Consequently, the bound for $\epsilon$ is conservative and requires the
structure of the network.
\end{rem}


Finally, we consider almost sure average consensus.  Note that the
gossip algorithm (\ref{eqa:alg_1}) can be viewed as a \emph{jump
linear system}, with i.i.d. system matrices $M(k),\ k\in
\mathbb{Z}_+$. For such systems, it is known (e.g.,
\cite[Corollary~3.46]{CoFrMa:04}) that almost sure convergence can
be implied from mean-square convergence. Therefore the result on
almost sure convergence is immediate.

\begin{cor}\label{cor:almsur}
Using the gossip algorithm (\ref{eqa:alg_1}) with the parameter
$\epsilon>0$ sufficiently small, the agents achieve almost sure
average consensus if and only if the digraph $\mathcal {G}$ is
strongly connected.
\end{cor}
\section{Special Topologies} \label{Sec5_SpeGra}

We turn now to a special class of topologies -- strongly connected
and balanced digraphs -- and investigate the required upper bound on
the parameter $\epsilon$ for the deterministic
algorithm~(\ref{eq:DetAlg}).  Furthermore, when these digraphs are
restricted to symmetric or cyclic respectively, we derive less
conservative $\epsilon$ bounds compared to the general one in
(\ref{eq:perturb_bound}).

Given a digraph $\mathcal {G} = (\mathcal {V}, \mathcal {E})$, its
\emph{degree} $d$ is defined by $d:=\max_{i \in \mathcal {V}}
\mbox{card}(\mathcal{N}^+_i)$.  In the deterministic
algorithm~(\ref{eq:DetAlg}) choose the updating and sending weights
to be respectively $a_{ij}=1/(2dn)$ and $b_{ij}=1/(dn)$, for every
$(j,i) \in \mathcal {E}$. This choice renders the two matrices
$I-2L$ and $S$ identical, when the digraph $\mathcal {G}$ is
balanced.  We will see that the equality $I-2L=S$ supports a
similarity transformation in dealing with the cyclic case below.

\begin{lem} \label{lem:balanced}
Suppose that the parameter $\epsilon$ satisfies $\epsilon \in
(0,2)$, and the zeros of the following polynomial for every $\mu
\neq 0$ with $|\mu - 1/(2n)|\leq1/(2n)$ lie strictly inside the unit
circle:
\begin{align} \label{eq:poly_complex}
p(\lambda) := \lambda^2 + \alpha_1\lambda + \alpha_0,
\end{align}
where $\alpha_0:=2\mu^2-3\mu-\epsilon+1$,
$\alpha_1:=3\mu+\epsilon-2$. Then the deterministic
algorithm~(\ref{eq:DetAlg}) achieves average consensus on strongly
connected and balanced digraphs.
\end{lem}
\noindent \emph{Proof.} We analyze the spectral properties of the
matrix $M$ in terms of those of the Laplacian matrix $L$. Let
$\mu_i$, $i=1,\ldots,n$, be the $i$th eigenvalue of $L$.  Since
$\mathcal {G}$ is balanced and all the updating weights are
$a_{ij}=1/(2dn)$, it follows from the Gershgorin Theorem
\cite[Chapter~6]{HorJoh:90} that $|\mu_i - 1/(2n)|\leq1/(2n)$.
Further, as $\mathcal {G}$ is strongly connected, by the
Perron-Frobenius Theorem \cite[Chapter~8]{HorJoh:90} we get that
$\mu_1=0$ is simple. Now substituting the equality $S=I-2L$ into
(\ref{eq:DetAlg}) one obtains
\begin{align*}
M=
\begin{bmatrix}
  I-L &  \epsilon I\\
  L &  I-2L-\epsilon I
\end{bmatrix}.
\end{align*}
Consider the characteristic polynomial of $M$:
\begin{align*}
\det(\lambda I - M) &= \det \left(
\begin{bmatrix}
  (\lambda-1)I+L &  -\epsilon I\\
  -L &  (\lambda-1+\epsilon)I + 2L
\end{bmatrix} \right) \\
&= \det\left( ((\lambda-1)I+L)((\lambda-1+\epsilon)I + 2L) - \epsilon L\right)\\
&= \det\left( (\lambda-1)(\lambda-1+\epsilon)I + 3(\lambda-1)L +
2L^2 \right).
\end{align*}
Here the second equality is due to that $(\lambda-1)I+L$ and $-L$
commute \cite{Sil:00}.  By spectral mapping one derives that the
$2n$ eigenvalues of $M$ can be obtained by solving the following $n$
equations:
\begin{align} \label{eq:eig_relation}
(\lambda-1)(\lambda-1+\epsilon) + 3(\lambda-1)\mu_i + 2\mu_i^2 = 0,
\ \ i=1,\ldots,n.
\end{align}
For $\mu_1=0$ we have from (\ref{eq:eig_relation}) that
$\lambda_1=1$ and $\lambda_2=1-\epsilon$. Since $\epsilon \in
(0,2)$, $\lambda_2 \in (-1,1)$. Now fix $i\in[2,n]$ so that $\mu_i
\neq 0$ and $|\mu_i - 1/(2n)|\leq1/(2n)$.  Note that the left hand
side of (\ref{eq:eig_relation}) can be arranged into the polynomial
$p(\lambda)$ in (\ref{eq:poly_complex}), whose zeros are inside the
unit circle. It follows that $1$ is a simple eigenvalue of $M$, and
all other eigenvalues have moduli smaller than one.  Therefore, by
Proposition~\ref{prop:Mean_AveCon} we conclude that average
consensus is achieved. \hfill $\blacksquare$

\begin{figure}[!t]
  \centering
  \includegraphics[width=0.73\textwidth]{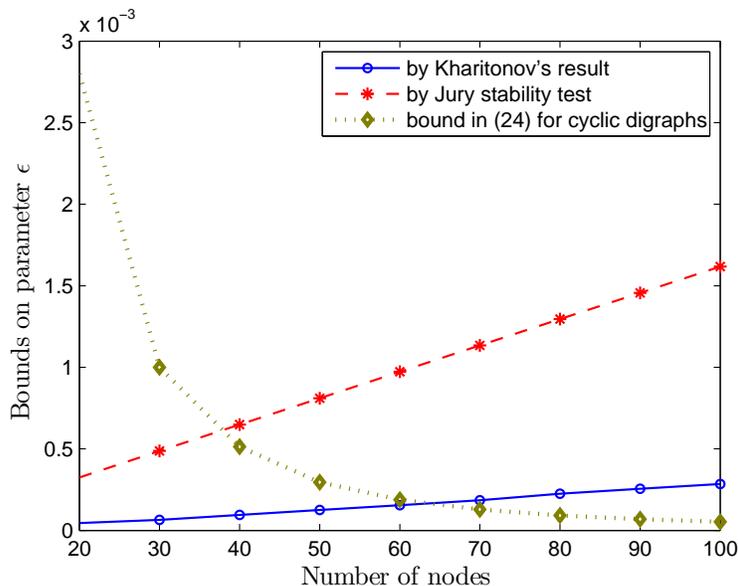}
  \vspace{-0.5cm}
  \caption{Upper bounds on parameter $\epsilon$ such that deterministic
  algorithm~(\ref{eq:DetAlg}) achieves average consensus on general strongly connected
 balanced digraphs (solid and dashed curves) and cyclic digraphs (dotted curve).} \vspace{0.5cm}
  \label{fig:Jury_vs_Kha}
\end{figure}

Now we investigate the values of $\epsilon$ that ensure the zeros of
the polynomial $p(\lambda)$ in (\ref{eq:poly_complex}) inside the
unit circle, which in turn guarantee average consensus on strongly
connected and balanced digraphs by Lemma~\ref{lem:balanced}. For
this, we view the polynomial $p(\lambda)$ as \emph{interval
polynomials} \cite{Barm:94} by letting $\mu$ take any value in the
square: $0 \leq \mbox{ Re}(\mu) \leq 1/n$, $-1/(2n) \leq \mbox{
Im}(\mu) \leq 1/(2n)$. Applying the bilinear transformation we
obtain a new family of interval polynomials:
\begin{align} \label{eq:bilinear}
\tilde{p}(\gamma) := (\gamma-1)^2 p\left( \frac{\gamma+1}{\gamma-1}
\right) =
(1+\alpha_0+\alpha_1)\gamma^2+(2-2\alpha_0)\gamma+(1+\alpha_0-\alpha_1).
\end{align}
Then by Kharitonov's result for the complex-coefficient case, the
stability of $\tilde{p}(\gamma)$ (its zeros have negative real
parts) is equivalent to the stability of eight extreme polynomials
\cite[Section~6.9]{Barm:94}, which in turn suffices to guarantee
that the zeros of $p(\lambda)$ lie strictly inside the unit circle.
Checking the stability of eight extreme polynomials results in upper
bounds on $\epsilon$ in terms of $n$. This is displayed in
Fig.~\ref{fig:Jury_vs_Kha} as the solid curve.  We see that the
bounds grow linearly, which is in contrast with the general bound
$\bar{\epsilon}$ in Proposition~\ref{prop:perturb_bound} that decays
exponentially and is known to be conservative. This is due to that,
from the robust control viewpoint, the uncertainty of $\mu$ in the
polynomial coefficients becomes smaller as $n$ increases.

Alternatively, we employ the Jury stability test \cite{Jury:88} to
derive that the zeros of the polynomial $p(\lambda)$ are strictly
inside the unit circle if and only if
\begin{align} \label{eq:jury}
\beta_0 := \left| \begin{array}{cc}
 1 & \alpha_0 \\
 \bar{\alpha}_0 & 1
\end{array} \right|>0,\ \ \ \beta_1 := {\small \left|
\begin{array}{cc}
\left|\begin{array}{cc}
 1 & \alpha_0 \\
 \bar{\alpha}_0 & 1
\end{array}\right| & \left|\begin{array}{cc}
 1 & \alpha_1 \\
 \bar{\alpha}_0 & \bar{\alpha}_1
\end{array}\right| \\ & \\
\left|\begin{array}{cc}
 1 & \bar{\alpha}_1 \\
 \alpha_0 & \alpha_1
\end{array}\right| & \left|\begin{array}{cc}
 1 & \bar{\alpha}_0 \\
 \alpha_0 & 1
\end{array}\right|
\end{array}\right|} >0.
\end{align}
Here $\beta_0$ and $\beta_1$ turn out to be polynomials in
$\epsilon$ of second and fourth order, respectively; the
corresponding coefficients are functions of $\mu$ and $n$. Thus
selecting $\mu$ such that $\mu \neq 0$ and $|\mu -
1/(2n)|\leq1/(2n)$, we can solve the inequalities in (\ref{eq:jury})
for $\epsilon$ in terms of $n$.  Thereby we obtain the dashed curve
in Fig.~\ref{fig:Jury_vs_Kha}, each plotted point being the minimum
value of $\epsilon$ over $1000$ random samples such that the
inequalities in (\ref{eq:jury}) hold.  This simulation confirms that
the true bound on $\epsilon$ for the zeros of $p(\lambda)$ to be
inside the unit circle is between the solid and dashed curves.
Since the discrepancy of these two curves is relatively small, it is
suggested that our previous analysis based on Kharitonov's result may
not very conservative.

Here ends our discussion on $\epsilon$ bounds for arbitrary balanced
(and strongly connected) digraphs.  In the sequel, we will further
specialize the balanced digraph $\mathcal {G}$ to be symmetric or
cyclic, respectively, and provide analytic $\epsilon$ bounds less
conservative than (\ref{eq:perturb_bound}) for the general case. In
particular, the exponent $n$ is not involved.

\subsection{Connected Undirected Graphs} \label{ch3_sec5_sub1}

A digraph $\mathcal {G}=(\mathcal {V},\mathcal {E})$ is
\emph{symmetric} if $(j,i)\in \mathcal {E}$ implies $(i,j)\in
\mathcal {E}$. That is, $\mathcal {G}$ is undirected.

\begin{prop} \label{prop:regular}
Consider a general connected undirected graph $\mathcal {G}$. Then
the deterministic algorithm~(\ref{eq:DetAlg}) achieves average
consensus if the parameter $\epsilon$ satisfies $\epsilon \in
(0,(1-(1/n))(2-(1/n))$.
\end{prop}
\noindent {\em Proof.} The symmetry of the undirected graph
$\mathcal {G}$ results in the symmetry of its Laplacian matrix $L$.
So all the eigenvalues $\mu_i$ of $L$ are real, and satisfy
$\mu_1=0$ and $(\forall i \in [2,n])\ \mu_i \in (0,1/n]$ ($\mathcal
{G}$ is connected).  For $\mu_1=0$ we know from
(\ref{eq:eig_relation}) that $\lambda_1=1$, and $\lambda_2 \in
(-1,1)$ since $0< \epsilon < (1-(1/n))(2-(1/n)) < 2$. For $\mu_i \in
(0,1/n]$, $i\in[2,n]$, consider again the polynomial $p(\lambda)$ in
(\ref{eq:poly_complex}).  According to the Jury stability test for
real-coefficient case \cite{Jury:65}, the zeros of $p(\lambda)$ are
strictly inside the unit circle if and only if
\begin{align*}
1+\alpha_0+\alpha_1>0,\ \ \ 1+\alpha_0-\alpha_1>0,\ \ \
|\alpha_0|<1.
\end{align*}
Straightforward calculations show that these conditions hold
provided $\epsilon \in (0,(1-(1/n))(2-(1/n))$.  Hence, the matrix
$M$ has a simple eigenvalue $\lambda_1=1$ and all others
$\lambda_2,\ldots,\lambda_{2n}\in(0,1)$.  Therefore, by
Proposition~\ref{prop:Mean_AveCon} the deterministic
algorithm~(\ref{eq:DetAlg}) achieves average consensus. \hfill
$\blacksquare$

It is noted that for connected undirected graphs, the upper bound on
$\epsilon$ ensuring average consensus grows as $n$ increases. This
characteristic is in agreement with that of the bounds for the more
general class of balanced digraphs as we observed in
Fig.~\ref{fig:Jury_vs_Kha}.

\subsection{Cyclic Digraphs} \label{ch3_sec5_sub2}

A digraph $\mathcal {G} = (\mathcal {V}, \mathcal {E})$ is
\emph{cyclic} if $\mathcal {V}=\{1,\ldots,n\}$ and $\mathcal
{E}=\{(1,2),(2,3),\ldots,(n-1,n),(n,1)\}$.  So a cyclic digraph is
strongly connected.

\begin{prop} \label{prop:cyclic}
Suppose that the digraph $\mathcal {G}$ is cyclic. Then the
deterministic algorithm~(\ref{eq:DetAlg}) achieves average consensus
if the parameter $\epsilon$ satisfies
\begin{align} \label{eq:bound_cyclic}
\epsilon \in \left(0, \frac{\sqrt{2}}{3+\sqrt{5}}
\left(1-|\lambda_3|\right)\right),\ \ \mbox{with $\lambda_3$ as in
(\ref{eq:eig_M0})}.
\end{align}
Further, in this case $|\lambda_3|=\sqrt{1-(1/n)+(1/(2n^2))+(1/n)
(1-1/(2n)) \cos2\pi /n}$.
\end{prop}

Before providing the proof, we state a perturbation result, the Bauer--Fike Theorem, for
diagonalizable matrices (e.g., \cite[Section~6.3]{HorJoh:90}).
Recall that the matrix $M$ in (\ref{eq:DetAlg}) can be written as $M
= M_0 + \epsilon F$, with $M_0$ and $F$ in (\ref{eq:M_0_F}).
Throughout this subsection, write $\lambda_i(\epsilon)$ for the
eigenvalues of $M$, and $\lambda_i$ for those of $M_0$.

\begin{lem}\label{lem:Bauer-Fike}
Suppose that $M_0$ is diagonalizable; i.e., there exist a
nonsingular matrix $V \in \mathbb{C}^{2n \times 2n}$ and a diagonal
matrix $J=\mbox{diag}(\lambda_1, \ldots, \lambda_{2n})$ such that
$M_0=VJV^{-1}$.  If $\lambda(\epsilon)$ is an eigenvalue of $M$,
then there is an eigenvalue $\lambda_i$ of $M_0$, for some $i\in
[1,2n]$, such that $|\lambda(\epsilon)-\lambda_i| \leq ||V||_2\
||V^{-1}||_2\ ||\epsilon F||_2$.
\end{lem}

In other words, every eigenvalue of the perturbed matrix $M$ lies in
a circle centered at some eigenvalue of the unperturbed matrix $M_0$
of the radius $(||V||_2\ ||V^{-1}||_2\ ||\epsilon F||_2)$. We now
present the proof of Proposition~\ref{prop:cyclic}.

\noindent {\em Proof of Proposition~\ref{prop:cyclic}.} Since the
digraph $\mathcal {G}$ is cyclic, we derive that its Laplacian
matrix $L$ is given by $L=\mbox{circ}(1/(2n),0,\ldots,0,-1/(2n))$ --
a \emph{circulant matrix} \cite{Davis:94} with the first row
$[1/(2n)\ 0\ \cdots\ 0\ -1/(2n)]\in \mathbb{R}^{1 \times n}$. Let
$\omega:=\mathrm{e}^{2\pi \iota/n}$ with $\iota:=\sqrt{-1}$. Then
the eigenvalues $\mu_i$ of $L$ are $\mu_i=(1/(2n))(1-\omega^{i-1})$,
$i=1,\ldots,n$. Rewrite the equation~(\ref{eq:eig_relation}) as
$(\lambda(\epsilon)-1)(\lambda(\epsilon)-1+\epsilon) +
3(\lambda(\epsilon)-1)\mu_i + 2\mu_i^2=0$. Then for $\mu_1=0$, we
have $\lambda_1(\epsilon)=1$ and $\lambda_2(\epsilon)=1-\epsilon$,
corresponding respectively to the eigenvalues $\lambda_1$,
$\lambda_2$ of $M_0$.  Evidently the upper bound in
(\ref{eq:bound_cyclic}) is strictly smaller than $2$; so
$\lambda_2(\epsilon) \in (-1,1)$.

We turn next to investigating the rest of the eigenvalues
$\lambda_3(\epsilon),\ldots,\lambda_{2n}(\epsilon)$, for which we
employ Lemma~\ref{lem:Bauer-Fike}.  Let $\Omega$ denote the $n
\times n$ \emph{Fourier matrix} given by
\begin{align*}
\Omega := \frac{1}{\sqrt{n}} \begin{bmatrix}
  1 & 1 & 1 & \cdots & 1 \\
  1 & \omega & \omega^2 & \cdots & \omega^{n-1} \\
  1 & \omega^2 & \omega^4 & \cdots & \omega^{2(n-1)} \\
  \vdots & \vdots & \vdots & \vdots & \vdots \\
  1 & \omega^{n-1} & \omega^{2(n-1)} & \cdots & \omega^{(n-1)(n-1)}
\end{bmatrix}.
\end{align*}
Note that $\Omega$ is unitary, i.e., $\Omega^{-1}=\Omega^*$ (the
conjugate transpose of $\Omega$).  It is a fact that every circulant
matrix can be (unitarily) diagonalized by $\Omega$
\cite[Theorem~3.2.1]{Davis:94}.  Now let $V :=\begin{bmatrix}
  \Omega & 0 \\
  \Omega & \Omega
\end{bmatrix}$, and consider
\begin{align*}
V^{-1} M_0 V = \begin{bmatrix}
  \Omega^* & 0 \\
  -\Omega^* & \Omega^*
\end{bmatrix} \begin{bmatrix}
  I-L & 0 \\
  L & S
\end{bmatrix} \begin{bmatrix}
  \Omega & 0 \\
  \Omega & \Omega
\end{bmatrix} = \begin{bmatrix}
  \Omega^*(I-L)\Omega & 0 \\
  0 & \Omega^*S\Omega
\end{bmatrix}.
\end{align*}
The last equality is due to $S=I-2L$. Hence $M_0$ is diagonalizable
via $V$, and its spectrum is
\begin{align*}
\sigma(M_0)=\sigma(I-L) \cup
\sigma(S)=\left\{1-\frac{1}{2n}(1-\omega^{i-1}),
1-\frac{1}{n}(1-\omega^{i-1}):i=1,\ldots,n\right\}.
\end{align*}
Also, by a direct calculation we get
$||V||_2=||V^{-1}||_2=\sqrt{(3+\sqrt{5})/2}$ and $||F||_2=\sqrt{2}$.
It then follows from Lemma~\ref{lem:Bauer-Fike} that for every
eigenvalue $\lambda_l(\epsilon)$ of $M$ there is an eigenvalue
$\lambda_{l'}$ of $M_0$, $l,l'\in [3,2n]$, such that
$|\lambda_l(\epsilon)-\lambda_{l'}| \leq ||V||_2\ ||V^{-1}||_2\
||\epsilon F||_2=\left((3+\sqrt{5})/2\right) \sqrt{2}\ \epsilon$. So
the upper bound of $\epsilon$ in (\ref{eq:bound_cyclic}) guarantees
$|\lambda_l(\epsilon)-\lambda_{l'}|<1-|\lambda_3|$; namely, the
perturbed eigenvalues still lie within the unit circle. Summarizing
the above we have $\lambda_1(\epsilon)=1$ and
$|\lambda_2(\epsilon)|,|\lambda_3(\epsilon)|,\ldots,|\lambda_{2n}(\epsilon)|
< 1$; therefore, the deterministic algorithm~(\ref{eq:DetAlg})
achieves average consensus by Proposition~\ref{prop:Mean_AveCon}.
Further, one computes that
\begin{align*}
|\lambda_3|&=\max_{i \in [2,n]}
\left\{\Big|1-\frac{1}{2n}(1-\omega^{i-1})\Big|,
\Big|1-\frac{1}{n}(1-\omega^{i-1})\Big|\right\}\\
&=\Big|1-\frac{1}{2n}+\frac{1}{2n}
\omega\Big|=\sqrt{1-\frac{1}{n}+\frac{1}{2n^2}+\frac{1}{n}
\left(1-\frac{1}{2n}\right) \cos \frac{2\pi}{n}}.
\end{align*} \hfill $\blacksquare$

Finally, in Fig.~\ref{fig:Jury_vs_Kha} we plot the upper bound on
$\epsilon$ in (\ref{eq:bound_cyclic}) for the class of cyclic
digraphs. We see that this bound decays as the number $n$ of nodes
increases, which contrasts with the bound characteristic of the more
general class of balanced digraphs.  This may indicate the
conservativeness of our current approach based on perturbation
theory.  Nevertheless, since the perturbation result used here is
specific only to diagonalizable matrices, the derived upper bound in
(\ref{eq:bound_cyclic}) is less conservative than the general one in
(\ref{eq:perturb_bound}).


\section{Numerical Examples} \label{Sec6_Simul}


\subsection{Convergence Paths}

\begin{figure}[!t]
  \centering
  \includegraphics[width=0.6\textwidth]{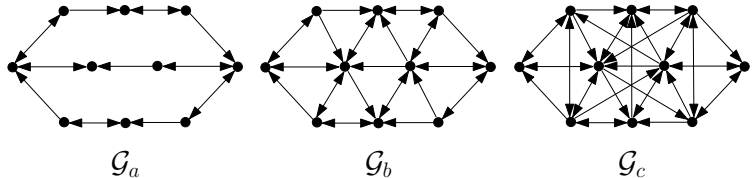}
  \vspace{-0.5cm}
  \caption{Three examples of strongly connected but non-balanced digraphs.}
  \label{fig:simu_10nodes}
\end{figure}

Consider the three digraphs displayed in
Fig.~\ref{fig:simu_10nodes}, with $10$ nodes and respectively $17$,
$29$, and $38$ edges.  Note that all the digraphs are strongly
connected, and in the case of uniform weights they are non-balanced
(indeed, no single node is balanced).  We apply both the
deterministic algorithm (\ref{eq:DetAlg}), with uniform weights
$a=1/(2 \mbox{card}(\mathcal {E}))$ and $b=1/\mbox{card}(\mathcal
{E})$, and the gossip algorithm (\ref{eqa:alg_1}), with uniform
weight $w=1/2$ and probability $p=1/\mbox{card}(\mathcal {E})$.

\begin{table}[!t]
\renewcommand{\arraystretch}{1.3}
\caption{Convergence factors $\lambda_2^{(d)}$ and $\lambda_2^{(g)}$
with respect to different values of parameter $\epsilon$.}
\label{tab:asymfac} \centering
\begin{tabular}{|c|cc|cc|cc|}
  \hline
                 & \multicolumn{2}{|c|}{$\epsilon=0.2$} & \multicolumn{2}{|c|}{$\epsilon=0.7$} & \multicolumn{2}{|c|}{$\epsilon=2.15$} \\
  \hline
                 & $\lambda_2^{(d)}$ & $\lambda_2^{(g)}$ & $\lambda_2^{(d)}$ & $\lambda_2^{(g)}$ & $\lambda_2^{(d)}$ & $\lambda_2^{(g)}$ \\
  $\mathcal {G}_a$ & $0.9963$ & $0.9963$ & $0.9993$ & $1.0003$ & $1.0003$ & $1.0020$ \\
  $\mathcal {G}_b$ & $0.9951$ & $0.9951$ & $0.9969$ & $0.9969$ & $0.9985$ & $1.0000$ \\
  $\mathcal {G}_c$ & $0.9883$ & $0.9883$ & $0.9930$ & $0.9930$ & $0.9966$ & $0.9993$ \\
  \hline
\end{tabular}
\end{table}

\begin{figure}[!t]
  \centering
  \includegraphics[width=0.55\textwidth]{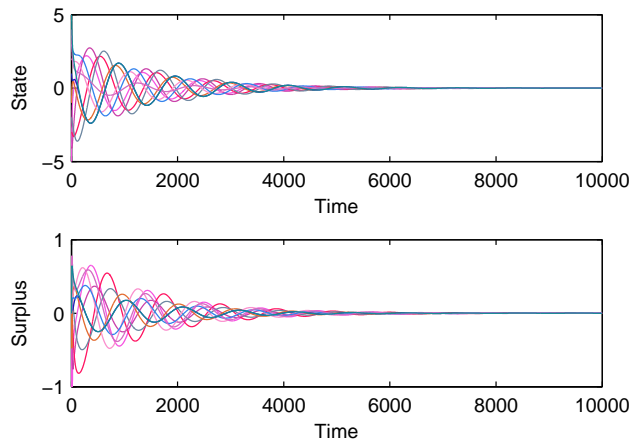}
  \vspace{-0.5cm}
  \caption{Convergence paths of states and surpluses: Obtained by applying the deterministic algorithm
  (\ref{eq:DetAlg}) with parameter $\epsilon=0.7$ on digraph $\mathcal {G}_a$.}
  \label{fig:Det_statesurplus}
\end{figure}

\begin{figure}[!t]
  \centering
  \includegraphics[width=0.55\textwidth]{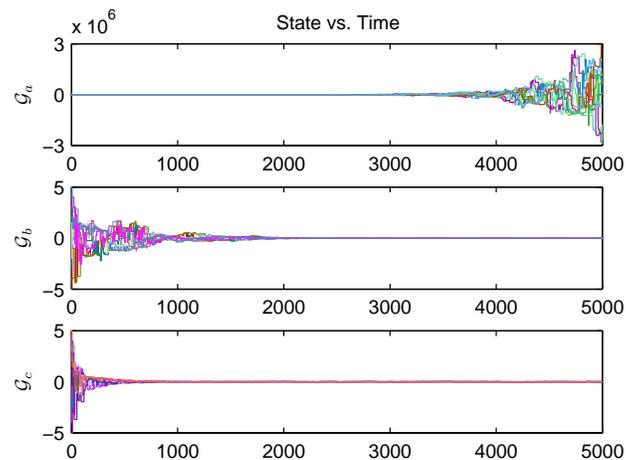}
  \vspace{-0.5cm}
  \caption{Sample paths of states: Obtained by applying the gossip algorithm (\ref{eq:DetAlg}) with parameter $\epsilon=0.7$ on digraphs $\mathcal {G}_a$, $\mathcal {G}_b$, and $\mathcal {G}_c$.}
  \label{fig:Gos_statecompare}
\end{figure}

The convergence factors $\lambda_2^{(d)}$ and $\lambda_2^{(g)}$ (see
Remarks~\ref{rem:confac_d} and \ref{rem:confac_g}) for three
different values of the parameter $\epsilon$ are summarized in
Table~\ref{tab:asymfac}.  We see that small $\epsilon$ ensures
convergence of both algorithms (the gossip algorithm
(\ref{eqa:alg_1}) requires smaller values of $\epsilon$ for
mean-square convergence), whereas large $\epsilon$ can lead to
instability.  Moreover, in those converging cases the factors
$\lambda_2^{(d)}$ and $\lambda_2^{(g)}$ decrease as the number of
edges increases from $\mathcal {G}_a$ to $\mathcal {G}_c$, which
indicates faster convergence when there are more communication
channels available for information exchange. We also see that the
algorithms are more robust on digraphs with more edges, in the sense
that a larger range of values of $\epsilon$ is allowed.

For a random initial state $x(0)$ with the average $x_a=0$ and the
initial surplus $s(0)=0$, we display in
Fig.~\ref{fig:Det_statesurplus} the trajectories of both states and
surpluses when the deterministic algorithm (\ref{eq:DetAlg}) is
applied on digraph $\mathcal {G}_a$ with parameter $\epsilon=0.7$.
Observe that asymptotically, state averaging is achieved and surplus
vanishes.  Under the same conditions, the gossip algorithm
(\ref{eqa:alg_1}), however, fails to converge, as shown in
Fig.~\ref{fig:Gos_statecompare}.  Applying algorithm
(\ref{eqa:alg_1}) instead on the digraphs $\mathcal {G}_b$ and
$\mathcal {G}_c$ which have more edges, average consensus is again
reached, and faster convergence occurs in $\mathcal {G}_c$ (see
Fig.~\ref{fig:Gos_statecompare}).


\subsection{Convergence Speed versus Parameter $\epsilon$}

\begin{figure}[!t]
  \centering
  \includegraphics[width=0.58\textwidth]{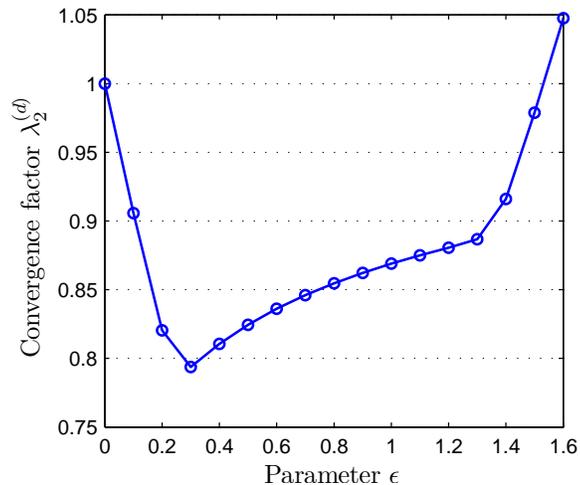}
  \vspace{-0.5cm}
  \caption{Convergence factor $\lambda_2^{(d)}$ of the deterministic algorithm
  (\ref{eq:DetAlg}) with respect to parameter $\epsilon$.}
  \label{fig:SpeedConv_det}
\end{figure}

\begin{figure}[!t]
  \centering
  \includegraphics[width=0.58\textwidth]{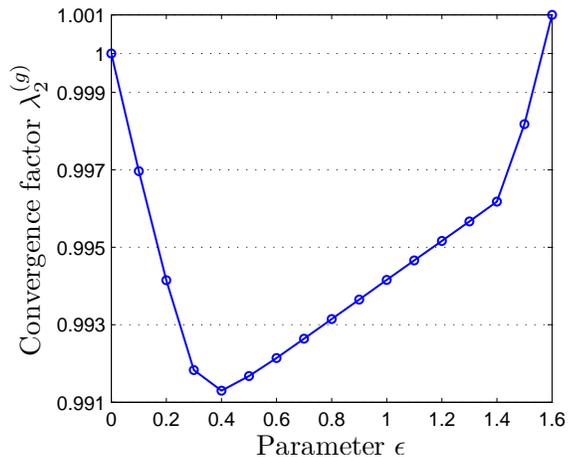}
  \vspace{-0.5cm}
  \caption{Convergence factor $\lambda_2^{(g)}$ of the gossip algorithm
  (\ref{eqa:alg_1}) with respect to parameter $\epsilon$.}
  \label{fig:SpeedConv_gos}
\end{figure}

%

We have seen that a sufficiently small parameter $\epsilon$ ensures
convergence of both algorithms (\ref{eq:DetAlg}) and
(\ref{eqa:alg_1}).  Now we investigate the influence of $\epsilon$
on the speed of convergence, specifically the convergence factors
$\lambda_2^{(d)}$ and $\lambda_2^{(g)}$.  To reduce the effect of
network topology in this investigation, we employ a type of random
digraphs where an edge between every pair of nodes can exist with
probability $1/2$, independent across the network and invariant over
time; we take only those that are strongly connected.

For the deterministic algorithm (\ref{eq:DetAlg}), consider random
digraphs of $50$ nodes and uniform weights $a=b=1/50$.
Fig.~\ref{fig:SpeedConv_det} displays the curve of convergence
factor $\lambda_2^{(d)}$ with respect to the parameter $\epsilon$,
each plotted point being the mean value of $\lambda_2^{(d)}$ over
$100$ random digraphs.  To account for the trend of this curve,
first recall from the perturbation argument for
Theorem~\ref{thm:DetAlg} that the matrix $M$ in (\ref{eq:DetAlg})
has two (maximum) eigenvalues $1$ when $\epsilon=0$, and small
$\epsilon$ causes that one of them (denote its modulus by
$\lambda_{in}$) moves into the unit circle.  Meanwhile, some other
eigenvalues of $M$ inside the unit circle move outward; denote the
maximum modulus among these by $\lambda_{out}$.  In our simulations
it is observed that when $\epsilon$ is small, $\lambda_2^{(d)} =
\lambda_{in}\ (> \lambda_{out})$ and $\lambda_{in}$ moves further
inside as perturbation becomes larger; so $\lambda_2^{(d)}$
decreases (faster convergence) as $\epsilon$ increases in the
beginning. Since the eigenvalues move continuously, there exists
some $\epsilon$ such that $\lambda_{in} = \lambda_{out}$,
corresponding to the fastest convergence speed. After that,
$\lambda_2^{(d)} = \lambda_{out}\ (> \lambda_{in})$ and
$\lambda_{out}$ moves further outside as $\epsilon$ increases; hence
$\lambda_2^{(d)}$ increases and convergence becomes slower, and
finally divergence occurs.


An analogous experiment is conducted for the gossip algorithm
(\ref{eqa:alg_1}), with random digraphs of $30$ nodes, uniform
probability $p=1/\mbox{card}(\mathcal {E})$, and uniform weight
$w_{ij}=1/2$.  We see in Fig.~\ref{fig:SpeedConv_gos} a similar
trend of $\lambda_2^{(g)}$ as the parameter $\epsilon$ increases,
though it should be noted that the changes in $\lambda_2^{(g)}$ are
smaller than those in $\lambda_2^{(d)}$.  From these observations,
it would be of ample interest to exploit the values of $\epsilon$
when the convergence factors achieve their minima, as well as the
upper bounds of $\epsilon$ ensuring convergence.
\section{Conclusions} \label{Sec7_Concl}


We have proposed distributed algorithms which enable networks of
agents to achieve average consensus on arbitrary strongly connected
digraphs. Specifically, in synchronous networks a deterministic
algorithm ensures asymptotic state averaging, and in asynchronous
networks a gossip algorithm guarantees average consensus in the
mean-square sense and with probability one.  To emphasize, our
derived graphical condition is more general than those previously
reported in the literature, in the sense that it does not require
balanced network structure; also, the matrix perturbation theory
plays an important role in the convergence analysis. Moreover,
special regular digraphs are investigated to give less conservative
bounds on the parameter $\epsilon$; and numerical examples are
provided to illustrate the convergence results, with emphasis on
convergence speed.


For future research, one direction of interest would be to extend
the deterministic algorithm (\ref{eq:DetAlg}) to the more realistic
scenario of \emph{switching} digraphs $\mathcal {G}(k) = (\mathcal
{V}, \mathcal {E}(k))$; namely, the network topology is
time-varying. If every $\mathcal {G}(k)$, $k \geq 0$, is strongly
connected, then it is possible to ensure convergence by introducing
\emph{slow switching} (e.g., dwell time) as in
\cite{RenBea:08,CaBeSw:08}. Under the weaker graphical condition
that digraphs $\mathcal {G}(k)$ are \emph{jointly} strongly
connected (\hspace{-0.01cm}\cite{JadLinMor:03,Lin:Mono:08}),
to verify if average consensus can be achieved seems to be more
challenging and requires further investigation.

On the other hand, in the literature on gossip algorithms \cite{BoGhPrSh:06,KaBaSr:07,FraFag:10},
a variety of practical communication issues have
been discussed such as link failure, message
collision, broadcast protocol, and synchronized node selection
(i.e., multiple nodes are selected at the same time).
We thus aim at addressing
these issues by making suitable extensions of our gossip algorithm
(\ref{eqa:alg_1}).


%
\IEEEpeerreviewmaketitle

%
%
%
%
%
\section*{Acknowledgment}

The authors would like to thank Sandro Zampieri for the helpful discussion,
and the anonymous reviewers for the valuable comments.

\ifCLASSOPTIONcaptionsoff
  \newpage
\fi



\bibliographystyle{IEEEtran}
\bibliography{DistributedControl,SwitchedSystem,EigPurterb}
\end{document}